\documentclass[12pt, square]{article}
\usepackage{latexsym,epsfig,graphicx,amsmath,amssymb,amscd,undertilde,multirow,psfrag,paralist,dsfont}
\usepackage[titletoc]{appendix}
\newcommand{\thetavec}{{\boldsymbol{\theta}}}
\newcommand{\veps}{\varepsilon}
\newcommand{\vepsvec}{{\boldsymbol{\varepsilon}}}
\newcommand{\Sigmavec}{{\boldsymbol{\Sigma}}}
\newcommand{\sigmavec}{{\boldsymbol{\sigma}}}
\newcommand{\wvec}{{\boldsymbol{w}}}
\newcommand{\yvec}{{\boldsymbol{y}}}

\newcommand{\zerovec}{{\boldsymbol{0}}}

\newcommand{\Ivec}{{\boldsymbol{I}}}
\newcommand{\Indfun}{{\mathds{1}}}
\newcommand{\betavec}{{\boldsymbol{\beta}}}

\newcommand{\E}{\mathbf{E}}

\newcommand{\pr}{{\rm Pr}}

\newcommand{\NOR}{{\rm N}}

\newcommand{\thetavechat}{\widehat{\thetavec}}

\newcommand{\sev}{\text{sev}}

\newcommand{\wh}{\widehat}

\newcommand{\Xvec}{\boldsymbol{X}}
\newcommand{\Zvec}{\boldsymbol{Z}}
\newcommand{\xvec}{\boldsymbol{x}}
\newcommand{\betavechat}{\widehat{\betavec}}
\newcommand{\D}{\mathcal{D}}

\newcommand{\betahat}{{\widehat{\beta}}}

\newcommand{\deltavec}{\boldsymbol{\delta}}
\newcommand{\tvec}{\boldsymbol{t}}
\newcommand{\qvec}{\boldsymbol{q}}

\newcommand{\SF}{{\mathcal{F}}}
\newcommand{\tm}{{t^{-}}}

\newcommand{\N}{{\textrm{N}}}

\newcommand{\ultimatestress}{\sigma_{\textrm{ult}}}

\begin{document}

\title{Statistical Analysis of Modern Reliability Data}
\author{Yueyao Wang\\
Department of Statistics\\
Virginia Tech\\
Blacksburg, VA 24061\\
\and
I-Chen Lee\\
Department of Statistics\\
National Cheng Kung University\\
Tainan, Taiwan\\
\and
Lu Lu\\
Department of Mathematics \& Statistics\\
University of South Florida\\
Tampa, FL,  33620\\
\and
Yili Hong\\
Department of Statistics\\
Virginia Tech\\
Blacksburg, VA 24061\\
}

\date{}

\maketitle
\begin{abstract}
Traditional reliability analysis has been using time to event data, degradation data, and recurrent event data, while the associated covariates tend to be simple and constant over time. Over the past years, we have witnessed the rapid development of sensor and wireless technology, which enables us to track how the product has been used and under which environmental conditions it has been used. Nowadays, we are able to collect richer information on covariates which provides opportunities for better reliability predictions. In this chapter, we first review recent development on statistical methods for reliability analysis. We then focus on introducing several specific methods that were developed for different types of reliability data with covariate information. Illustrations of those methods are also provided using examples from industry. Test planning is also an important part of reliability analysis. In addition to data analysis, we also provide a briefly review on recent developments of test planning and then focus on illustrating the sequential Bayesian design with an example of fatigue testing for polymer composites. The chapter is concluded with some discussions and remarks.

\end{abstract}
\newpage


\section{Introduction}
\subsection{Background}
Traditional reliability data analysis mainly use time to event data, degradation data, and recurrent event data to make reliability predictions~\cite{meekerescobar1998}. The covariate information involved in the reliability analysis is usually time-invariant and the number of covariates is typically small. For time to event data, parametric models such as the Weibull and lognormal distributions are popular and accelerated failure time models are often used to incorporate covariate information on accelerating factors. For degradation data, the general path models and stochastic models are the common choices and the covariate information is often incorporated through regression type of models. The recurrent event data are often modeled by the event intensity models or mean cumulative functions with regression type of models that are often used to incorporate covariates.

With technological advances, new information on covariates become available. Products and systems can be equipped with sensors and smart chips to keep track of various information on the field usage of product units, number of transfers, and environmental conditions such as temperature and humidity. Such covariate information often change over time, so we refer to them as dynamic covariate information. Because the dynamic covariates often come in large volume and variety, it presents big data opportunities and challenges in the area of reliability analysis (e.g., \cite{MeekerHong2014} and \cite{HongZhangMeeker2018}). Dynamic covariate data can be used for modeling and prediction of reliability because units under heavy usage often fail sooner than those lightly used. In recent years, more statistical methods for dynamic covariates have been being developed to make use of this new type of covariate data.

Another important area of reliability analysis is about test planning, which focuses on how to efficiently collect various types of data to make better prediction of reliability. For accelerated life tests (ALTs), it is especially challenging to timely collect sufficient failure data because the data collection is a time-consuming process and often requires using expensive equipment for testing units under elevated stress conditions. In some laboratories, there are typically one or two machines available for testing certain material. In this case, it is impractical to test multiple samples simultaneously and therefore limits the total obtainable sample size. Another challenge with traditional test planning is that it typically relies on a single set of best guess of the parameter values, which may lead to suboptimal designs when the specified parameter values are not accurate. Due to these challenges, sequential designs become popular where earlier test results can be utilized to determine the test conditions for later runs. In addition, Bayesian methods can be used to leverage prior information from the expert's knowledge or related historical data to inform the test planning. The objective of this chapter is to review current development and then introduce the statistical methods for dynamic covariates and sequential Bayesian design (SBD) for ALT.

\subsection{Related Literature}
In lifetime data analysis, product usage information has been used to improve reliability model. Lawless et al.~\cite{LawlessCrowderLee2009} consider warranty prediction problem using product usage information on return units. Constant usage information are used in \cite{Guoetal2009} and \cite{LuAndersonCook2010}. Averaged product use-rate information are used in \cite{HongMeeker2010}. Nelson~\cite{nelson2001} and Voiculescu et al.~\cite{Voiculescuetal2007} use cumulative exposure model in ALT and reliability analysis. Hong and Meeker~\cite{HongMeeker2013} use cumulative exposure model to incorporate dynamic covariates and apply it to the Product D2 application.

In degradation data analysis stochastic process models are widely used. The Wiener process (\cite{Whitmore1995,DoksumHoyland1992, Wang2010}), Gamma process (\cite{LawlessCrowder2004}), and Inverse Gaussian process (\cite{WangXu2010,YeChen2014}) are among popular models in this class. The general path models are also widely used, which include \cite{LuMeeker1993,MeekerEscobarLu1998,BagdonaviciusNikulin2001,BaeKuoKvam2007,Duanetal2017}. For accelerated destructive degradation tests, the typical work includes \cite{EscobarMeekerKuglerKramer2003,Xieetal2018,Dingetal2019}. Hong et al.~\cite{HongDuanetal2015} and Xu et al.~\cite{XuHongJin2015} develop degradation model using the general path model framework to incorporate dynamic covariate information.

For recurrent events data, the nonhomogeneous Poisson process (NHPP) and the renewal process (RP) are widely used (e.g., \cite{ZhaoLiu2003,HongLiOsborn2015}). Kijima~\cite{Kijima1989} introduce virtual age models which can model imperfect repairs.  Pham and Wang~\cite{PhamandWang1996} develop a quasi-renewal process, and Doyen and Gaudoin~\cite{DoyenandGaudoin2004} propose models for imperfect repairs. The trend-renewal process (TRP) proposed in \cite{LindqvistElvebakkHeggland2003} can include the NHPP and RP as special cases, which has been used in \cite{YangShi2012,PietznerWienke2013, Yangetal2017} and other places. Xu et al.~\cite{Xuetal2017} develop a multi-level trend renewal process (MTRP) model for recurrent event with dynamic covariates.

For test planning, the optimum designs in traditional test planning framework are developed using non-Bayesian approaches (e.g., \cite{meeker1984,nelson1990}) and the true parameters are assumed to be known. Bayesian test planning for life data is developed in \cite{zhang2005, zhang2006, Hongetal2015}. King et al.~\cite{Kingetal2016} develop optimum test plans for fatigue test of polymer composites. Lee et al.~\cite{lee2018} develop SBD test planning for polymer composites and Lu et al.~\cite{LuLeeHong2019} extend it to test planning with dual objectives.

\subsection{Overview}
The rest of this chapter is organized as follows. Section~\ref{sec:time.to.event} describes an application on time-to-event data with dynamic covariates. Section~\ref{sec:deg.data} illustrates the modeling of degradation with dynamic covariates. Section~\ref{sec:recurent.event} describes the MTRP model for describing recurrent event data with dynamic covariates. Section~\ref{sec:seq.alt} introduces SBD strategies for ALTs. Section~\ref{sec:concluding.remark} contains some concluding remarks.



\section{Time to Event Data Analysis}\label{sec:time.to.event}
In this section, we briefly introduce the application of using dynamic covariates for time to event prediction as described in Hong and Meeker~\cite{HongMeeker2013}.

\subsection{Background and Data}
\begin{figure}
	\begin{center}
		\begin{tabular}{cc}
			\includegraphics[width=.47\textwidth]{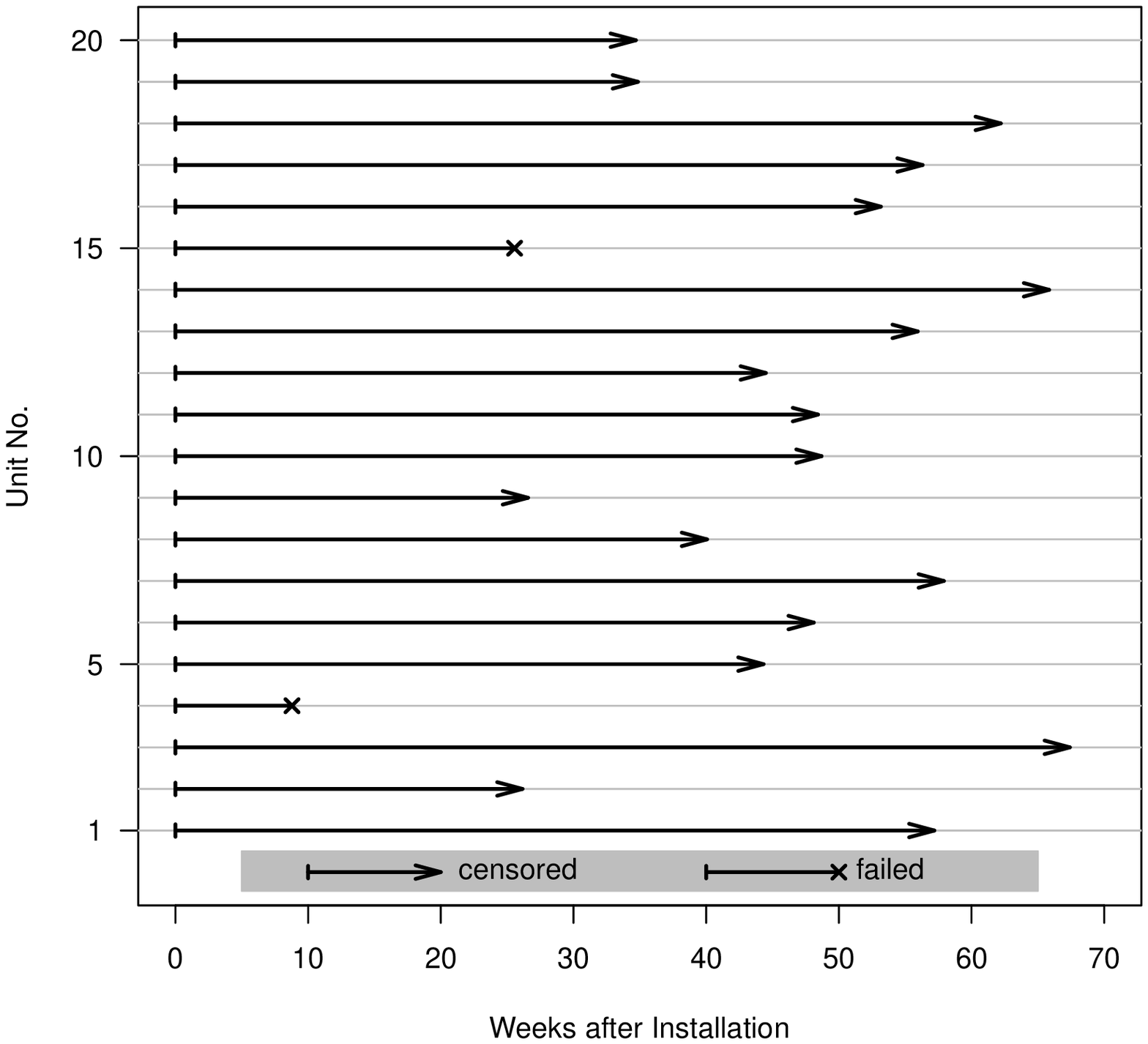}
			&
			\includegraphics[width=.47\textwidth]{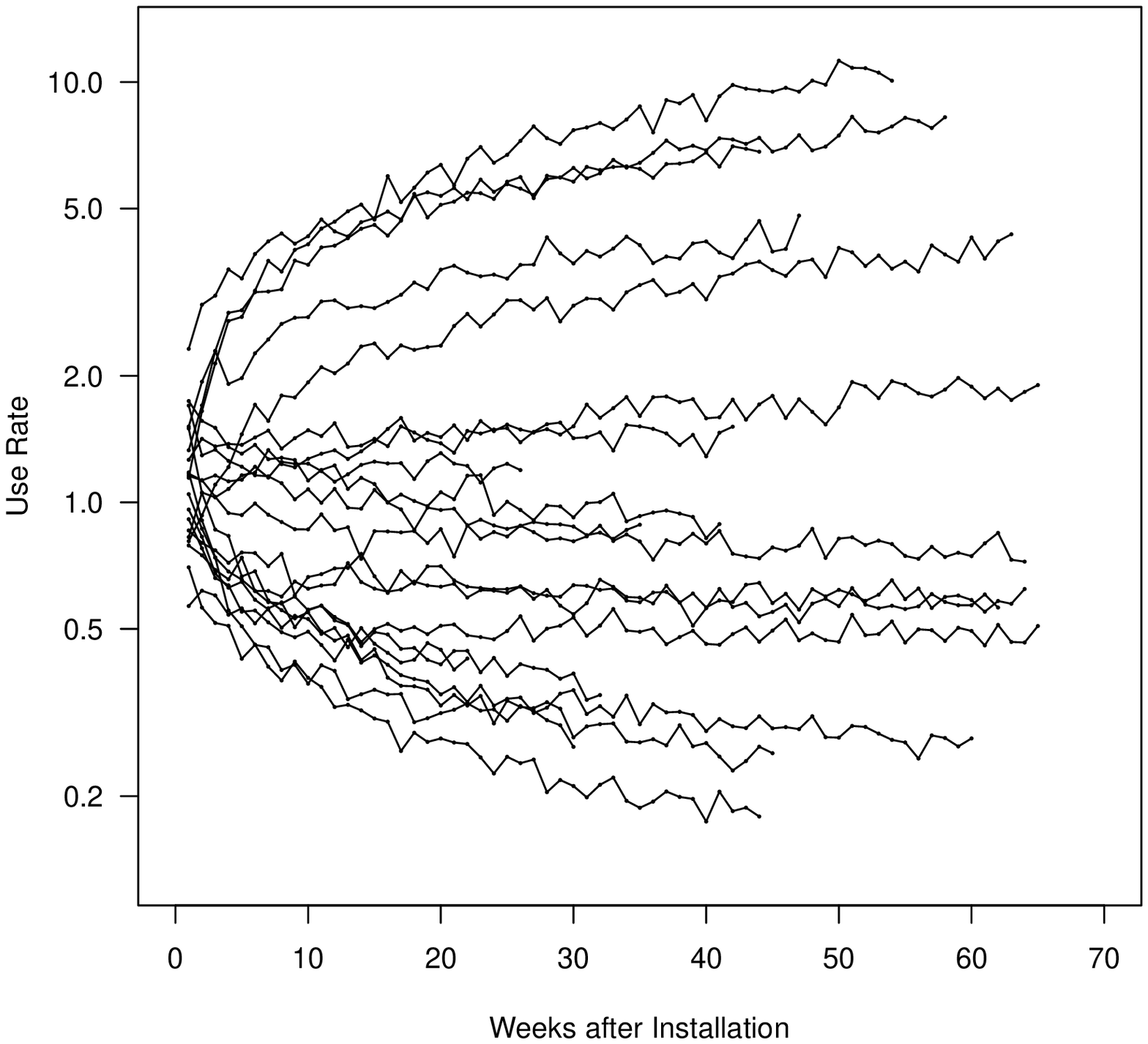}\\
			{\small (a) Failure-time Data}& {\small (b) Use-rate processes}
		\end{tabular}
	\end{center}
	\caption{(a) The event plot for a subset of the Product D2 failure-time data and (b) the corresponding plot of the use-rate trajectories. \emph{Figure reproduced with permission}.}\label{fig:Prod.D2.event.plot}
\end{figure}

A general method was developed by Hong and Meeker \cite{HongMeeker2013} to model failure-time data with dynamic covariates. The work was motivated by the Product D2 application, which is a machine used in office/residence. Product D2 is similar to high-end copy machine where the number of pages used could be recorded dynamically. For this product, the use-rate data $R(t)$ (cycles/week) were collected weekly as a time series for those units connected to the network. This information could be downloaded automatically in addition to the failure-time data. In the Product D2 dataset, data were observed within a 70-week period and 69 out of 1800 units failed during the study period. Figure~\ref{fig:Prod.D2.event.plot} illustrates the event plot of the failure-time data and the use-rate over time for a subset of the data.

\subsection{Model for Time to Event and Parameter Estimation}
Three sets of observable random variables: the failure time, censoring indicator and dynamic covariate over time are are considered, which are denoted by $\{\,T,\Delta,\Xvec(T)\,\}$. The observed data are described by $\{\,t_i,\delta_i,\xvec_i(t_i)\,\}$. Here $n$ denotes the number of units in the dataset, $t_i$ is the failure time or time in service, and $\delta_i$ is the observed censoring indicator (i.e., it equals to 1 if unit $i$ fails and 0 otherwise). The $\xvec_i(t_i)=\left\{x_i(s): 0\leq s\leq t_i\right\}$ is the observed dynamic covariate information of unit $i$ from the time $0$ to $t_i$, where $x_i(s)$ is the observed covariate value at time $s$ for unit $i$. Particularly for Product D2, we use $X(t) = \log[R(t)/R_0(t)]$ as the form of the covariate in the model, where $R_0(t)$ is the baseline use-rate that is chosen to be a typical constant use rate.

The cumulative exposure model in \cite{bagnik2001} is used to model the failure-time data with dynamic covariate. The cumulative exposure $u(t)$ is defined as,
$$u(t)=u[t;\beta,\xvec(t)]=\int_{0}^t\exp[\beta x(s)]ds,$$
where $\beta$ represents the influence of the covariate on the exposure. When the cumulative exposure of a unit reaches a random threshold $U$ at time $T$, the unit fails. This establishes a relationship between $U$ and $T$, that is,
\begin{align}\label{eqn:cumulative.exposure.model}
U=u(T)=\int_{0}^T\exp[\beta x(s)]ds.
\end{align}
Under the above model and the covariate history $\xvec(\infty)$, the cumulative distribution function (cdf) of the failure time $T$ is
\begin{align*}
F(t;\beta,\thetavec_0)&=\pr(T\leq t)=\Pr\{U\leq u[t;\beta,\xvec(t)]\}=F_0\left\{u[t;\beta,\xvec(t)];\thetavec_0\right\}
\end{align*}
and probability density function (pdf) is $f(t;\beta,\thetavec_0)=\exp[\beta x(t)]f_0\left\{u[t;\beta,\xvec(t)];\thetavec_0\right\}.$
Here $\thetavec_0$ is the parameter in the baseline cdf of the cumulative exposure threshold $U$ and $f_0(u;\thetavec_0)$ is the pdf of $U$. In the Product D2 application, the baseline cumulative exposure distribution $F_0(u;\thetavec_0)$ was modeled by the Weibull distribution, of which the cdf and pdf are
\begin{align*}
F_0(u;\thetavec_0)=\Phi_{\sev}\left[\frac{\log(u)-\mu_0}{\sigma_0}\right] \text{ and }
f_0(u;\thetavec_0)=\frac{1}{\sigma_0 u}\phi_{\sev}\left[\frac{\log(u)-\mu_0}{\sigma_0}\right].
\end{align*}
In the above expression, $\thetavec_0=(\mu_0,\sigma_0)'$, where $\mu_0$ and $\sigma_0$ are the location and scale parameters. Also, $\Phi_{\sev}(z)=1-\exp[-\exp(z)]$, and $\phi_{\sev}(z)=\exp[z-\exp(z)]$. Lognormal and other log-location-scale distributions can also be used if they are considered appropriate for certain applications.

\subsection{Model for Covariates}

To model the covariate process, we use the linear mixed effect model. In particular, $X(t)$ is modeled as
\begin{align}\label{eqn:model.xt}
X_i(t_{ij})=\eta+Z_i(t_{ij})\wvec_i+\veps_{ij}.
\end{align}
In model \eqref{eqn:model.xt}, $\eta$ is the constant mean, and the term $Z_i(t_{ij})\wvec_i$ is used to model variation at individual level. Here $Z_i(t_{ij})=\left[1,\,\, \log(t_{ij})\right]$ and $\wvec_i$ is the vector of random effects of the initial covariate at time 0 and the changing rate for unit $i$. It is assumed that $\wvec_i=(w_{0i},w_{1i})'\sim \NOR(\zerovec,\Sigmavec_{\wvec})$ with the covariance matrix
$$\Sigmavec_{\wvec}=\left(
\begin{array}{cc}
\sigma_{1}^2&\rho\sigma_{1}\sigma_{2}\\
\rho\sigma_{1}\sigma_{2}&\sigma_2^2\\
\end{array}\right),$$
and  $\veps_{ij}\sim \NOR(0,\sigma^2)$ is the error term.

The parameter estimation is conducted in two parts since parameters in the failure-time model $\thetavec_T=(\thetavec_0',\beta)'$ and covariates process model $\thetavec_X = (\eta,\Sigmavec_{\wvec},\sigma^2)$ are separate, using the maximum Likelihood (ML) method. The joint likelihood for $\thetavec_T$ and $\thetavec_X$ is
\begin{equation} \label{eqn:likelihood}
L(\thetavec_T, \thetavec_X)=L(\thetavec_T)\times L(\thetavec_X).
\end{equation}
The first component of (\ref{eqn:likelihood}) is the likelihood function of the failure-time data, which is
\begin{align}\label{eqn:lik.failure.time}
&L(\thetavec_T)=\prod_{i=1}^{n}\left\{\exp[\beta x_i(t_i)]
f_0\left(u[t_i;\beta,\xvec_i(t_i)];\thetavec_0\right)\right\}^{\delta_i}
\left\{1-F_0\left(u[t_i;\beta,\xvec_i(t_i)];\thetavec_0\right)\right\}^{1-\delta_i}.
\end{align}
The second component of (\ref{eqn:likelihood}) is the likelihood of covariate data, which is
\begin{align}
&L(\thetavec_X)=\prod_{i=1}^n\int_{\wvec_i}\left\{\prod_{t_{ij}\leq t_i}f_{1}\left[x_i(t_{ij})-\eta-Z_i(t_{ij})\wvec_i;\sigma^2\,\right]\right\}f_{2}(\wvec_i;\,\Sigmavec_{\wvec})d\wvec_i.
\end{align}
In the above equation, $f_{1}(\,\cdot\,)$ is the pdf of a univariate normal and $f_{2}(\,\cdot\,)$ is the pdf of a bivariate normal distribution.

\subsection{Reliability Prediction}

In order to predict future field failures, the distribution of the remaining life (DRL) is considered in the prediction procedure. The DRL describes the distribution of $T_i$ for unit $i$ given $T_i>t_i$ and $\Xvec_i(t_i)=\xvec_i(t_i)$. Particularly, the probability of unit $i$ failing within the next $s$ time units given it has survived by the time $t_i$ is
\begin{align}
\rho_i(s;\thetavec)=\pr[t_i<T_i\leq t_i+s|T_i>t_i,\Xvec_i(t_i)],\quad
s>0,
\end{align}
where $\thetavec$ denotes all parameters. Then $\rho_i(s;\thetavec)$ can be further expressed as
\begin{align}\label{eqn:rho.individual}
\rho_i(s;\thetavec)&=\E_{\Xvec_i(t_i,t_i+s)|\Xvec_i(t_i)=\xvec_i(t_i)}\left\{\pr[t_i<T_i\leq
t_i+s|T_i>t_i,\Xvec_i(t_i),\Xvec_i(t_i,t_i+s)]\right\}\\[.8ex]\nonumber
&=\frac{\E_{\Xvec_i(t_i,t_i+s)|\Xvec_i(t_i)=\xvec_i(t_i)}\left\{F_0\left(u[t_i+s;\beta,\Xvec_i(t_i+s)];\thetavec_0\right)\right\}
	-F_0\left(u[t_i;\beta,\xvec_i(t_i)];\thetavec_0\right)}
{1-F_0\left(u[t_i;\beta,\xvec_i(t_i)];\thetavec_0\right)}
\end{align}
where $\Xvec_i(t_1,t_2)=\left\{X_i(s): t_1<s\leq t_2\right\}$. Since model \eqref{eqn:model.xt} is assumed for $\Xvec_i(t_i,t_i+s)$ and $\Xvec_i(t_i)=\xvec_i(t_i)$, the multivariate normal distribution theory can be used to obtain the conditional distribution.

The Monte Carlo simulation is used to evaluate $\rho_i(s;\thetavechat)$ since an analytical expression for $\rho_i(s;\thetavec)$ is unavailable. The following procedure is used to compute $\rho_i(s;\thetavechat)$.

\begin{enumerate}
	\item Substitute $\thetavec_X$ with the ML estimates $\thetavechat_X$ in the distribution of $\Xvec_i(t_i,t_i+s)|\Xvec_i(t_i)=\xvec_i(t_i)$, and draw $\Xvec_i^{\ast}(t_i,t_i+s)$ from the distribution.
	\item Let $\Xvec_i^{\ast}(t_i+s)=\{\xvec_i(t_i),\Xvec_i^{\ast}(t_i,t_i+s)\}$ be the simulated covariate process in the time interval $\left(t_i, t_{i}+s\right)$.
	\item Compute the DRL given $\Xvec_i^{\ast}(t_i,t_i+s)$ and the ML estimates $\thetavechat_T$ of $\thetavec_T = (\thetavec_0', \beta)$ by
	\begin{align*}
	\rho_i^{\ast}(s;\thetavechat)&=\frac{F_0\left(u[t_i+s;\betahat,\Xvec_i^{\ast}(t_i+s)];\thetavechat_0\right)
		-F_0\left(u[t_i;\betahat,\xvec_i(t_i)];\thetavechat_0\right)}
	{1-F_0\left(u[t_i;\betahat,\xvec_i(t_i)];\thetavechat_0\right)}.
	\end{align*}
	\item Repeat steps 1-3 $M$ times and obtain $\rho_i^{\ast m}(s;\thetavechat), m=1,\cdots,M$.
	\item The estimate is computed by $\rho_i(s;\thetavechat)=M^{-1}\sum_{m=1}^M\rho_i^{\ast m}(s;\thetavechat).$
\end{enumerate}

The confidence intervals (CIs) for $\rho_i(s;\thetavechat)$ can be obtained through the following procedure:

\begin{enumerate}
	\item Draw $\thetavechat_T^{\ast\prime}$ and $\thetavechat_X^{\ast\prime})'$ from $\NOR(\thetavechat_T,\Sigma_{\thetavechat_T})$ and $\NOR(\thetavechat_X,\Sigma_{\thetavechat_X})$, respectively.
	\item Let $\thetavechat^{\ast}=(\thetavechat_T^{\ast\prime},\thetavechat_X^{\ast\prime})'$ and obtain $\rho_i^{\ast\ast}(s;\thetavechat^{\ast})$ following the above algorithm.
	\item Repeat steps 1-2 $B$ times to obtain $\rho_i^{\ast\ast b}(s)=\rho_i^{\ast\ast b}(s;\thetavechat^{\ast b}), b=1,\cdots, B$.
	\item The $100(1-\alpha)\%$ CI is computed by $\left[\rho_i^{\ast\ast [\alpha B]}(s),\,\, \rho_i^{\ast\ast [(1-\alpha)B]}(s)\right]$. Here $\rho_i^{\ast\ast [b]}(s)$ is the $[b]$th ordered value of $\rho_i^{\ast\ast b}(s)$ and $[\,\cdot\,]$ is the function for rounding to the nearest integer.
\end{enumerate}

\begin{figure}
	\begin{center}
		\includegraphics[width=.5\textwidth]{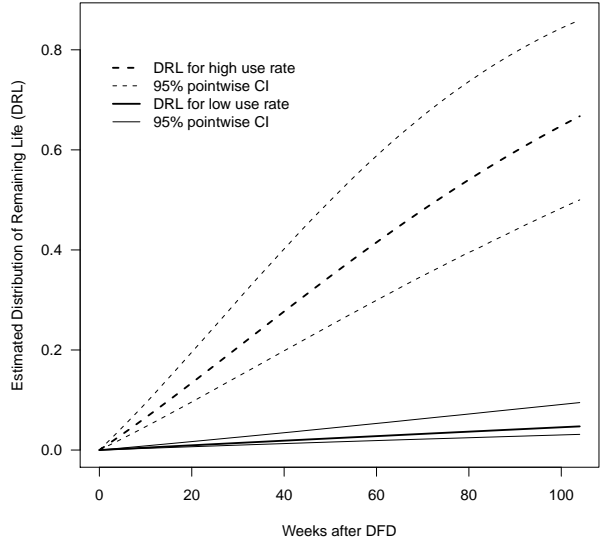}
	\end{center}
	\caption{The estimated DRLs and the 95\% pointwise CIs for two representative at-risk units. \emph{Figure reproduced with permission}.}\label{fig:ind.rhoi.ci}
\end{figure}

Figure~\ref{fig:ind.rhoi.ci} shows the estimated DRL for two representative units. One unit has a higher use rate which increases quickly over time $(\wh{w}_0=0.4061,\wh{w}_1=0.4184)$ and the other has a lower use rate which increases slowly over time $(\wh{w}_0=0.1704,\wh{w}_1=0.0168)$. The trends in the plot are consistent with our expectation that units with higher use rates tend to have higher failure risk.

To assess the prediction variability, one may also want to calculate the prediction interval (PI) of individual remaining life, denoted by $\left[\utilde{S}_i,\,\widetilde{S}_i\,\right]$. A $100(1-\alpha)\%$ PI of the remaining lifetime can be obtained by using the method introduced by \cite{LawlessFredette2005} as in
\begin{eqnarray}\label{eqn:cali.ind}
\rho_i(\utilde{S}_i;\wh\thetavec)=v_{\alpha/2}\quad \text{ and }
\quad \rho_i(\widetilde{S}_i;\wh\thetavec)=v_{1-\alpha/2}.
\end{eqnarray}
Here $v_{\alpha}$ is the $\alpha$ quantile of the $\rho_i(S_i;\wh\thetavec)$ distribution, and $S_i$ represents the remaining life for unit $i$. The $v_{\alpha}$ can be obtained through a Monte Carlo simulation.

In real applications, obtaining the predicted cumulative number of failures is also important because this could help with the decisions for warranty cost control or the long-term production plan. Suppose $N(s)=\sum_{i\in RS}I_i(s)$ is the total failure counts at time $s$ after DFD. The $RS$ is the risk set at DFD in this expression and $I_i(s)$ is the binary indicator for the occurrence of a failure at time $s$ with $I_i(s)\sim\text{Bernoulli}[\rho_i(s;\thetavec)].$ Let $F_{N}(n_k;\thetavec), n_k=0,1,\dots, n^\ast$ denote the cdf of $N(s)$, where $n^\ast$ is the count of units in the risk set. Then $F_{N}(n_k;\thetavec)$ can be computed in its explicit form using a discrete Fourier transformation \cite{Hong2013}. The PI for $N(s)$ can be calculated similarly as the individual predictions.

\begin{figure}
	\begin{center}
		\includegraphics[width=.5\textwidth]{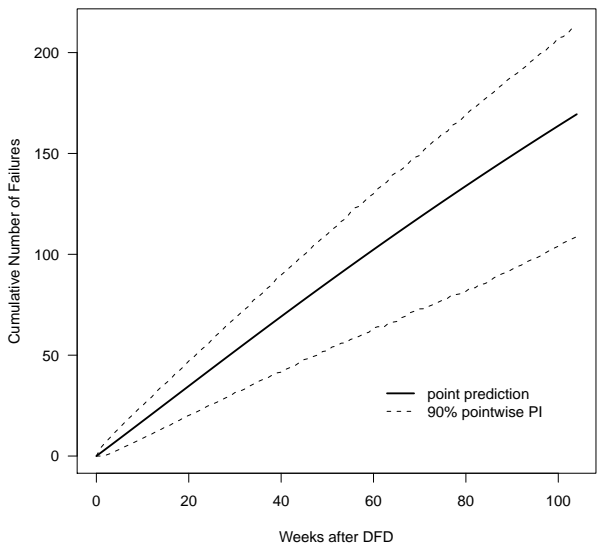}
	\end{center}
	\caption{The point predictions and the pointwise 90\% PIs for the cumulative number of failures after the DFD out of the 1731 units at risk. \emph{Figure reproduced with permission}.}\label{fig:cali.pi}
\end{figure}

For the Product D2 application, 1731 units are remained at risk after 69 out of 1800 installed units failed. The point predictions and the pointwise 90\% PIs of the total number of failures after the DFD are shown in Figure~\ref{fig:cali.pi}.



\section{Degradation Data Analysis}\label{sec:deg.data}
In this section, we briefly introduce how to leverage the dynamic covariates for modeling degradation data as described in Hong et al.~\cite{HongDuanetal2015}.

\subsection{Background and Data}

\begin{figure}
\centering
\begin{tabular}{cc}
\includegraphics[width=.4\textwidth]{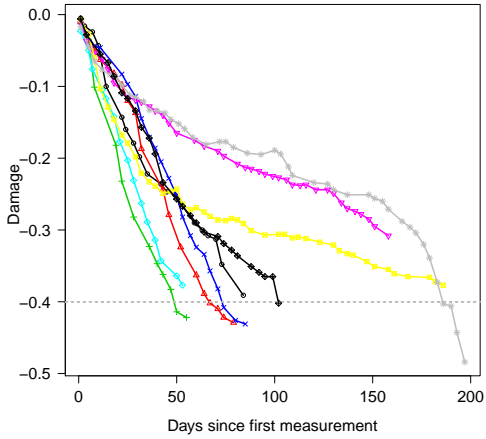} &
\includegraphics[width=.4\textwidth]{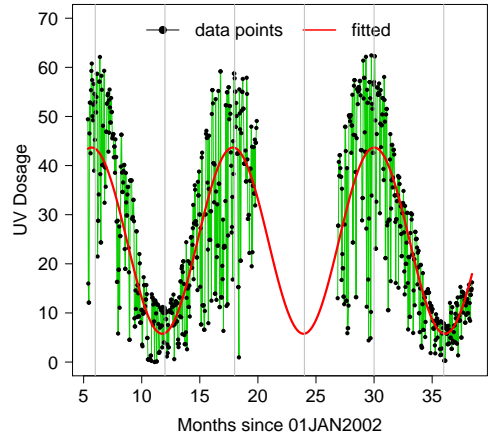}\\
(a) Degradation paths & (b) Daily UV dosage\\
\includegraphics[width=.4\textwidth]{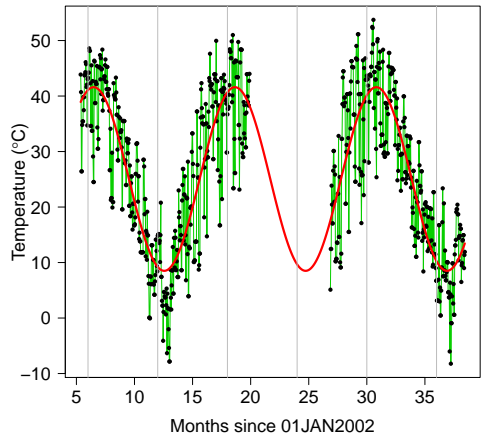} &
\includegraphics[width=.4\textwidth]{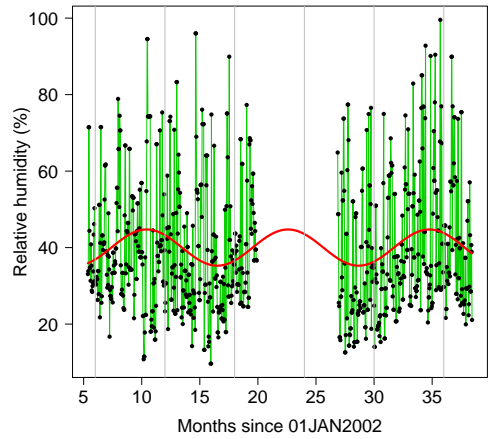}\\
(c) Daily temperature & (d) Daily RH\\
\end{tabular}
\caption{Plots of (a) nine representative degradation paths and (b)-(d) dynamic covariate information on the daily UV dosage, temperature and relative humidity for a single sample. The black dots connected by green lines show the daily values. The vertical lines show the time windows by every 6 months from January 2002. The red smooth curves are the estimated mean process. \emph{Figure reproduced with permission}.}\label{fig:deg-path}
\end{figure}

Hong et al.~\cite{HongDuanetal2015} develop a general path model utilizing shape-restricted splines with random effects to model the degradation process with dynamic covariates. This paper considers an application of modeling the photodegradation process of organic coating materials due to exposure to the time-varying ultraviolet (UV) radiation and the outdoor environmental conditions. In this work, to study the service life of a coating material, scientists at NIST placed 36 specimens in outdoor setting with varied UV spectrum and intensity, temperature, and relative humidity (RH) recorded over an approximate 5-year period. The specimens started at different time points to allow different degradation paths to be observed. For each specimen, degradation measurements were taken periodically using Fourier transform infrared (FTIR) spectroscopy. Since a particular compound or structure is tied with a peak at a certain wavelength on the FTIR spectrum, the change in the height of the peak was used to measure the decrease in the concentration of the compound. One of the compounds of interest for the NIST data was C-O stretching of aryl ether, which was measured at the wavelength 1250 cm. Figure \ref{fig:deg-path}(a) shows the degradation paths of nine representative specimens with varied starting times in the study. We can observe very different trajectories with the degradation rate varies over time and among different specimens as well. Figures \ref{fig:deg-path}(b)-(d) show the dynamic covariate information on the daily UV dosage, RH, and temperature as well as the fitted smooth lines for showing the mean process of one specimen over the study period. The vertical lines are used to label time windows separated by every six months. We can observe both a seasonal pattern and a random oscillation of the daily records for each individual covariate. There are stronger seasonal patterns for the UV dosage and temperature than the RH. There also appears to be a larger variation of the daily observations in the summer than in the winter, which indicates a varied degree of variability of the covariates over different time periods.

\subsection{Model for Degradation Paths and Parameter Estimation}

The general additive model for degradation data with dynamic covariates is given in the form
\begin{align}\label{eqn:gen.model}
y_{i}(t_{ij})&=D[t_{ij};\xvec_i(t_{ij})]+G(t_{ij};\mathbf{w}_i)+\veps_{i}(t_{ij}),
\end{align}
where $y_{i}(t_{ij})$ for $i=1,\cdots,n, j=1,\cdots,n_i$ is the degradation measurement at time $t_{ij}$ for unit $i$, $\veps_{i}(t_{ij})\sim \N(0,\sigma_{\veps}^2)$ denotes the measurement error, and $\xvec_i(t_{ij})=[x_{i1}(t_{ij}), \dots,x_{ip}(t_{ij})]'$ is the vector containing the dynamic covariate information at the time $t_{ij}$.  The actual degradation level at $t_{ij}$ is modeled by $D[t_{ij};\xvec_i(t_{ij})]+G(t_{ij};\mathbf{w}_i)$ as the sum of a fixed component and a random component. The fixed component is the population common degradation path, modeled in a cumulative damage form given by
\begin{align}\label{eqn:spline.model}
D[t_{ij};\xvec_i(t_{ij})]&=\beta_0+\sum_{l=1}^{p}\int_0^{t_{ij}}f_{l}[x_{il}(u);\beta_l]du.
\end{align}
This model incorporates the dynamic covariates through the covariate-effect functions $f_l(\cdot)$ for $l=1,\cdots,p$. Here, $\beta_0$ is the initial degradation, $f_l[x_{il}(u);\beta_l]$ is the $l$th covariate-effect of $x_{il}(u)$  on the degradation process at time $u$, and $\int_{0}^{t_{ij}}f_l[x_{il}(u);\beta_l]du$ is the cumulative effect of $x_{il}$ up to time $t_{ij}$. The random component includes the random effect terms for modeling the unit-to-unit variation, which is specified in $G(t_{ij};\mathbf{w}_i)=w_{0i}+w_{1i}t_{ij}$. Here, $\mathbf{w}_i=(w_{0i},w_{1i})'$ is the vector of random effects for the initial degradation and the growth rate over time, and it is assumed to follow a bivariate normal distribution $\N(0, \Sigma_\mathbf{w})$ with the covariance matrix
$$\Sigma_\mathbf{w}=\left[
\begin{array}{cc}
\sigma_0^2&\rho\sigma_0\sigma_1\\
\rho\sigma_0\sigma_1&\sigma_1^2\\
\end{array}
\right].$$
Also we use $\sigmavec_\mathbf{w}=(\sigma_0,\sigma_1,\rho)'$ to denote all distinct parameters included in $\Sigma_\mathbf{w}$.

The ML method is used for estimating the parameters. Since the degradation measurements and the dynamic covariates are observed at discrete time points, the discrete version of the degradation path model is used for computing the likelihood by replacing $D[t_{ij};\xvec_i(t_{ij})]$ in (\ref{eqn:spline.model}) by
\begin{align}\label{eqn:spline.model.disc}
D[t_{ij};\xvec_{i}(t_{ij})]=\beta_0+\sum_{l=1}^p\sum_{u_{ik}\leq t_{ij}}f_{l}[x_{il}(u_{ik});\beta_l](u_{ik}-u_{i,k-1}),
\end{align}
where $u_{ik}$ is the $k$th time point when the degradation and covariates are measured for unit $i$ and $u_{i0}=0$. Let $\thetavec_{D} = \left\{ \betavec, \sigmavec_{w}, \sigma_{\veps} \right\}$ denote all the model parameters. Then the likelihood is
\begin{align}\label{eqn:lik.hood}
L(\thetavec_{D})=\prod_{i=1}^n\int_{\mathbf{w}_i}\left[\prod_{t_{ij}\leq t_{in_i}}\frac{1}{\sigma_{\veps}}\phi\left\{\frac{C[y_{i}(t_{ij});\xvec_{i}(t_{ij}),\mathbf{w}_i]}
{\sigma_{\veps}}\right\}g_{\mathbf{w}_i}(\mathbf{w}_i;\sigmavec_{w})\right]d\mathbf{w}_i
\end{align}
where $C[y_{i}(t_{ij});\xvec_{i}(t_{ij}),\mathbf{w}_i]=y_{i}(t_{ij})-D[t_{ij};\xvec_{i}(t_{ij})]-G(t_{ij};\mathbf{w}_i)$, $\phi(\cdot)$ and $g_{w_i}(\cdot)$ are the pdfs of a standard normal distribution and a bivariate $\N(0,\Sigma_{\mathbf{w}})$ distribution, respectively.

Considering there was not sufficient knowledge on what might be a sensible form for the covariate-effect functions, the paper chose to estimate the $f_l(\cdot)$ using a linear combination of spline bases. To leverage the physical understanding of the relationships between the degradation process and the covariates, the shape-restricted splines \cite{Meyer2008} were used to ensure monotonic decreasing bases (I-splines) for the UV dosage and temperature and concave bases (C-splines) for the RH. Let $B_{lq}[x_{il}(u_{ik})]$ for $q=1,\cdots, a_l$ denote the spline bases for the covariate $x_l$, then the covariate-effect function is modeled as
$$f_{l}[x_{il}(u_{ik});\beta_l]=\sum_{q=1}^{a_l}B_{lq}[x_{il}(u_{ik})]\beta_{lq},$$
where $\beta_{lq}$'s are the spline coefficients. Define $U_{lq}(t_{ij})=\sum_{u_{ik}\leq t_{ij}}B_{lq}[x_{il}(u_{ik})](u_{ik}-u_{i,k-1})$. Then the model in (\ref{eqn:gen.model}) with $D[t_{ij};\xvec_i(t_{ij})]$ given in (\ref{eqn:spline.model.disc}) can be written as a linear mixed effects model in the form of
$\yvec_i=\Xvec_i\betavec+\Zvec_iw_i+\vepsvec_i,$
where
$$\Xvec_i=\left[
\begin{array}{cccccccc}
1&U_{11}(t_{i1})&\cdots&U_{1a_1}(t_{i1})&\cdots&U_{p1}(t_{i1})&\cdots&U_{pa_p}(t_{i1})\\
\vdots&\vdots&\ddots&\vdots&\ddots&\vdots&\ddots&\vdots\\
1&U_{11}(t_{in_i})&\cdots&U_{1a_1}(t_{in_i})&\cdots&U_{p1}(t_{in_i})&\cdots&U_{pa_p}(t_{in_i})\\
\end{array}
\right],\quad
\Zvec_i=\left[
\begin{array}{cc}
1&t_{i1}\\
\vdots&\vdots\\
1&t_{in_i}\\
\end{array}
\right],
$$
and the coefficient vector $\betavec=(\betavec_u',\betavec_c')'$, where $\betavec_u$ and $\betavec_c$ denote the unconstrained and constrained parameters, respectively.

The following algorithm was proposed \cite{HongDuanetal2015} to obtain the ML estimate $\thetavechat_{D}$ that maximizes~(\ref{eqn:lik.hood}):
\begin{enumerate}
\item Initiallize $\sigmavec_\mathbf{w}$ and $\sigma_{\veps}$ by fitting a linear mixed-effects model with no constraints.

\item Compute $V_{i}=\Zvec_i\Sigma_{w}\Zvec_i'+\sigma_{\epsilon}^2\Ivec_i$.

\item The mixed primal-dual bases algorithm in \cite{FraserMassam1989} is used to estimate $\betavec$. That is to minimize $\sum_{i=1}^n(\yvec_i-\Xvec_i\betavec)'V_i^{-1}(\yvec_i-\Xvec_i\betavec)$ subject to $\betavec_c \geq \textbf{0}$.

\item Fit a linear mixed-effects model $r_i=\Zvec_i\mathbf{w}_i+\vepsvec_i$ with $r_i=\yvec_i-\Xvec_i\betavechat$ to get the updated estimates of $\sigmavec_\mathbf{w}$ and $\sigma_{\veps}$.

\item Repeat steps 2 to 4 until the estimated parameters converge.
\end{enumerate}

With the shape-restricted splines, the ML estimates of some parameters might locate on the boundary of the parameter space. In this case, the bootstrap method is useful for assessing the variability and making inference about the parameters. An adjusted bootstrap procedure by \cite{CarppenterGoldsteinrasbash2003} was applied to resample the residuals and the estimated random effects for constructing bootstrap resamples of the original data to avoid underestimating variability and producing too narrow CIs. Then the bias-corrected bootstrap CIs were constructed based on obtaining the ML estimates of model parameters using the above mentioned algorithm for a large number of bootstrap samples.

\subsection{Model for Covariates}

To predict the degradation process and reliability, it is necessary to model the dynamic covariate process through a parametric model. Hong et al.~\cite{HongDuanetal2015} propose the following model
$$X_j(t)=\text{Tr}_j(t)+\text{S}_j(t)+\xi_j(t),$$
where $\text{Tr}_j(t)$ models the long-term trend of the covariate process for the $j$th covariate, $\text{S}_j(t)$ captures the seasonal pattern, and $\xi_j(t)$ depicts the random error which is assumed to be a stationary process. For the NIST outdoor weathering data, there was no significant long-term trend observed, and hence $\text{Tr}_j(t)=\mu_j$ for $j=1,2,3$. However, the seasonal pattern was quite prominent and there were seasonal effects observed for both the mean and variance of the process. So two sine functions were included in both the seasonal and error terms (except for RH which shows no seasonal effect assumed for the variation of the process from Figure \ref{fig:deg-path}) in the following form
\begin{align}\label{eqn:mean.structure}
\left[
\begin{array}{c}
\text{S}_1(t)\\
\text{S}_2(t)\\
\text{S}_3(t)\\
\end{array}
\right]=
\left[
\begin{array}{c}
\kappa_1\sin\left[\frac{2\pi}{365}(t-\eta_1)\right]\\[.5ex]
\kappa_2\sin\left[\frac{2\pi}{365}(t-\eta_2)\right]\\[.5ex]
\kappa_3\sin\left[\frac{2\pi}{365}(t-\eta_3)\right]\\[.5ex]
\end{array}
\right], \quad
\left[
\begin{array}{c}
\xi_1(t)\\
\xi_2(t)\\
\xi_3(t)\\
\end{array}
\right]=
\left[
\begin{array}{c}
\left(1+\nu_1\left\{1+\sin\left[\frac{2\pi}{365}(t-\varsigma_1)\right]\right\}\right)\veps_1(t)\\[.5ex]
\left(1+\nu_2\left\{1+\sin\left[\frac{2\pi}{365}(t-\varsigma_2)\right]\right\}\right)\veps_2(t)\\[.5ex]
\veps_3(t)\\[.5ex]
\end{array}
\right].
\end{align}\\
To capture the autocorrelation within and among the covariate processes, a lag-2 VAR model [i.e. Var(2)] was used, where the error term was modeled by
\begin{align}\label{eqn:error.structure}
\left[
\begin{array}{c}
\veps_1(t)\\
\veps_2(t)\\
\veps_3(t)\\
\end{array}
\right]=
Q_1\left[
\begin{array}{c}
\veps_1(t-1)\\
\veps_2(t-1)\\
\veps_3(t-1)\\
\end{array}
\right]+
Q_2\left[
\begin{array}{c}
\veps_1(t-2)\\
\veps_2(t-2)\\
\veps_3(t-2)\\
\end{array}
\right]+\left[
\begin{array}{c}
e_1(t)\\
e_2(t)\\
e_3(t)\\
\end{array}
\right]
\end{align}
In the above equation, $Q_1$ and $Q_2$ are regression coefficients matrices, and $[e_1(t),e_2(t),e_3(t)]'\sim\textrm{N}(0,\Sigma_e)$ are multivariate normal random errors that do not change over time.

The parameters in models (\ref{eqn:mean.structure}) and (\ref{eqn:error.structure}) are estimated in two steps. First, the ML estimates of the seasonal effects in the process mean and variance structures are obtained by ignoring the autocorrelation in the error terms. Then the VAR model is fitted to the residuals calculated from the first step using the multivariate least squares approach \cite{Lutkepohl2005}. The bootstrap method is used for obtaining the CIs of the parameters in the dynamic covariate process.

\subsection{Reliability Prediction}

\begin{figure}
\begin{center}
\includegraphics[width=0.45\textwidth]{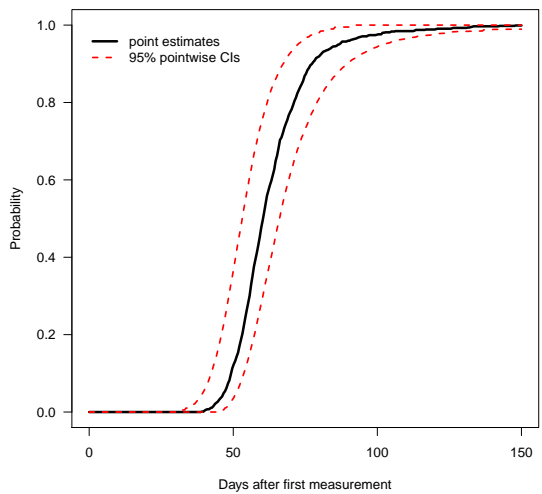}
\end{center}
\caption{The estimated cdf and corresponding 95\% pointwise CIs for a population of units with random starting time between 161 and 190 days. \emph{Figure reproduced with permission}.}\label{fig:rand.ind}
\end{figure}

To predict the failure time and reliability, let $\D_f$ denote the threshold for a soft failure. For any $\Xvec(\infty)=\xvec(\infty)$ and $\mathbf{w}$, the degradation path is fixed and the failure time can be determined by
\begin{align}\label{eqn:td.df}
t_{\D}=\min\{t: D[t;\xvec(\infty)]+G(t;\mathbf{w})=\D_f\}.
\end{align}
However, for a random unit, the covariate process $\Xvec(\infty)$ and $w$ are random. Hence, the cdf of the failure time, $T=T[\D_f,\Xvec(\infty),\mathbf{w}]$, can be defined as
\begin{align}\label{eqn:Ft}
F(t;\thetavec)=\E_{\Xvec(\infty)}\E_\mathbf{w}\pr\left\{T[\D_f,\Xvec(\infty),\mathbf{w}]\leq t\right\},\quad t>0,
\end{align}
with $\thetavec=\{\thetavec_{D},\thetavec_{X}\}$ denoting all the unknown parameters. There is usually no closed form expression of $F(t;\thetavec)$. Hence, the cdf at any estimated $\thetavechat$ is estimated through Monte Carlo simulation outlined in the following steps \cite{HongDuanetal2015}.
\begin{enumerate}
\item One need to simulate the covariate process based on the estimated parameter $\thetavechat_X$.
\item Then one can simulate the random effects $w$ from $\textrm{N}(0,\Sigma_{\mathbf{w}})$ with the estimated parameter $\thetavechat_D$.
\item Compute $D[t;\Xvec(\infty)]+G(t;\mathbf{w})$ based on the simulated covariate process and random effects.
\item For the degradation path in step 3, determine the failure-time $t_{\D}$ by Eqn.~(\ref{eqn:td.df}).
\item Repeat steps 1 to 4 for $M$ times to obtain the simulated failure-times $t_{\D}^m, m=1,\dots,M$. Then $F(t;\thetavechat)$ is estimated by the empirical cdf, $F(t;\thetavechat)=M^{-1}\sum_{m=1}^M\Indfun_{(t_{\D}^m\leq t)}$.
\end{enumerate}
By using the bootstrap approach, the point estimates and the CIs of $F(t;\thetavec)$ can be calculated using the sample mean and quantiles of the bootstrap version of the estimates of $F(t;\thetavechat)$ based on a large number of bootstrap estimates $\thetavechat$. By using $\D_f=-0.4$, $M=200$ Monte Carlo simulations, and $10000$ bootstrap samples, Figure \ref{fig:rand.ind} shows the predicted $F(t;\thetavec)$ and its 95\% pointwise CIs for the NIST coating degradation data. We can see that for a population of units with random starting time between 161 and 190 days, a majority of the population will fail between 50 to 150 days in service.

\section{Recurrent Event Data Analysis}\label{sec:recurent.event}
In this section, we briefly introduce the multi-level trend renewal process (MTRP) model and its application on the Vehicle B data as described in Xu et al.~\cite{Xuetal2017}.

\subsection{Background and Data}
\begin{figure}
\begin{center}
\begin{tabular}{cc}
\includegraphics[width=0.45\textwidth]{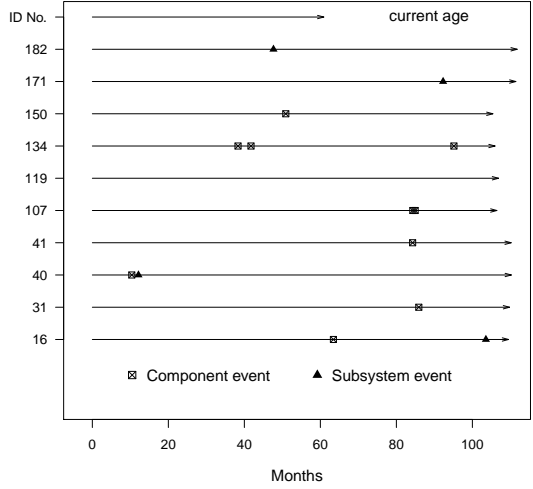} &
\includegraphics[width=0.45\textwidth]{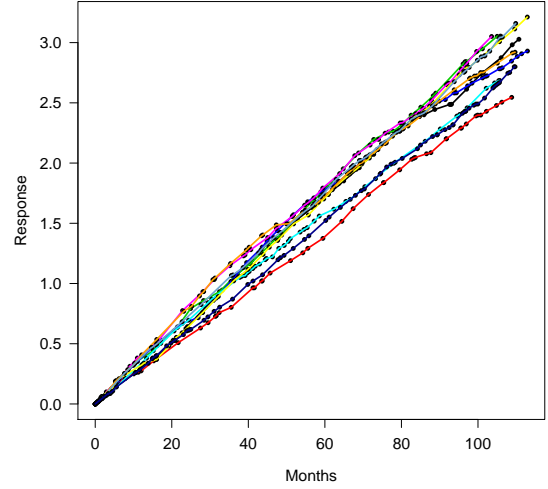}\\
(a) Recurrent event processes for    & (b) Cumulative usage processes for \\
 a subset of system units &  a subset of system units
\end{tabular}
\caption{Plots of (a) the event processes and (b) the cumulative usage processes for ten randomly selected units in the Vehicle B fleet. \emph{Figure reproduced with permission}.}\label{fig:SAeventdisplay}
\end{center}
\end{figure}

Xu et al.~\cite{Xuetal2017} consider the modeling and analysis of the Vehicle B data, which consist of recurrent event data from a batch of industrial systems. Vehicle B is a two-level repairable system. During its life span, Vehicle B may experience event at subsystem level (e.g., engine failures) and/or event at component level (e.g., oil pump failures). In the field data, we have $n=203$ units from a $110$-month observation period. There are 219 component events and 44 subsystem events observed during the study period. Figure~\ref{fig:SAeventdisplay}(a) shows the event plot for ten randomly selected units. We also have the cumulative usage information available for each unit, which is dynamic covariate. The cumulative usage information is shown in Figure~\ref{fig:SAeventdisplay}(b). The goal is to make a prediction for the cumulative number of component event occurrences at a future time.

We need some notation to introduce the MTRP model. Suppose there are $n$ units under observation from time 0 to $\tau_i$. Let $X_i(t)$ be the time-dependent covariate at time $t$ for system $i$. Let $N_{is}(t)$ be the number of subsystem events and and $N_{ic}(t)$  be the number of component events up to time $t$. The total number of replacement events is $N_i(t)=N_{is}(t)+N_{ic}(t)$. The subsystem event time is sorted as $0<t_{i1}^{s}<\cdots<t_{i,N_{is}(\tau_i)}^{s}<\tau_i$. The component event time is sorted as $0<t_{i1}^{c}<\cdots<t_{i,N_{ic}(\tau_i)}^{c}<\tau_i$. Let $0<t_{i1}<\cdots<t_{i,N_{i}(\tau_i)}<\tau_i$ be the replacement event times, regardless of the types.

\subsection{The MTRP Model and Parameter Estimation}\label{sec:repairable.model}
For a two-level repairable system, Xu et al.~\cite{Xuetal2017} propose the following MTRP model to describe events occurred at component level. In particular, the intensity function is
\begin{align}\label{eqn:component.intensity}
\lambda_i^c(t|\SF_{i,\tm}; \thetavec^{c})=h^c\left\{\Lambda_i^{s}(t|\SF_{i,\tm}^{\,s})-\Lambda_i^{ s}\left[t_{i, N_i(\tm)}\,|\,\SF_{i,t^-_{i,N_i(\tm)}}^{\,s}\right];\thetavec^{c}\right\}\lambda_i^{s}(t|\SF_{i,\tm}^{\,s};\thetavec^{c}).
\end{align}
Here $\SF_{i,\tm}^{\,s}$ denotes the historical information. In this multi-level model framework, the effect of subsystem events on the component event process is modeled by $\lambda_i^{s}(t|\SF_{i, \tm}^{\,s};\thetavec^{c})$, which takes the form
\begin{align} \label{eqn:trendftn}
\lambda_i^{s}(t|\SF_{i, \tm}^{\,s};\thetavec^{c})=h^{s}\{\Lambda_i(t)-\Lambda_i[t^s_{i,N_{is}(\tm)}] ; \thetavec^{c} \}\lambda_i(t; \thetavec^{c}),
\end{align}
Here, $\thetavec^{c}$ denotes the unknown parameters. The cumulative event intensity functions can be obtained as $\Lambda_i(t)=\int_0^{t}\lambda_i(u; \thetavec^c)\ du$, and $\Lambda_i^{s}(t|\SF_{i,\tm}^{\,s})=\int_{0}^{t}\lambda_i^{s}(u|\SF_{i,u^{-}}^{\,s}; \thetavec^c)\ du$. The baseline function $\lambda_i(t; \thetavec^{c})$ models the intensity of the component process when there is no event adjustment, and the function $h^s(\cdot)$ is used to model the adjustment for effect of events from the subsystem. The renewal distribution function $F^c(\cdot)$ is used to describe the distribution of gap times under the transformed scale. The model in \eqref{eqn:trendftn} can be extended to incorporate dynamic covariates and random effects.

To model the dynamic covariates, the intensity function can be extended as
\begin{align}\label{eqn:covariateintensity}
\lambda_i(t; \thetavec^c)=\lambda_b(t)\exp\{\gamma g[X_{i}(t)]\},
\end{align}
where $\lambda_b(t)$ denotes intensity trend function under the baseline and $\gamma$ is the regression coefficient. In the Vehicle B application, we use $g[X_i(t)]=\log[X_i(t)]$.  To incorporate random effects, the intensity function can be further extended as
\begin{align}\label{eqn:sysranintensity}
\lambda_i(t; \thetavec^c)=\lambda_b(t)\exp\{\gamma \log[X_{i}(t)]+ w_i\}.
\end{align}
Here $w_i$'s are independent and identically distributed with $\N(0, \sigma_r^2)$. The MTRP with random effects is referred to as $\textrm{HMTRP}(F^c, F^{s}, \lambda_i)$, in which the HMTRP stands for heterogenous MTRP.

To estimate the model parameters, one need to construct the likelihood function. The component events data can be denoted as $\{t_{ij}, \delta_{ij}^c\}$ with $t_{ij}$ be the event time and $\delta_{ij}^c$ be the component-event indicator. The event history is denoted as $\SF=\{N_{ic}(u), N_{is}(u), X_i(u): 0< u \le \tau_i,\ i=1, \cdots, n\}$. The likelihood function is
\begin{align} \label{eqn:comlikelihood}
L(\thetavec^c)=
=&\prod_{i=1}^n\prod_{j=1}^{N_i(\tau_i)+1}\left(\left\{f^c[\Lambda_i^{s}(t_{ij}|\SF_{i,t_{ij}^{-}}^{s})-\Lambda_i^{ s}(t_{i,j-1}|\SF_{i, t_{i,j-1}^{-}}^{s})]\lambda_i^{s}(t_{ij}|\SF_{i, t_{ij}^{-}}^{s}; \thetavec^c)\right\}^{\delta_{ij}^c}\right. \nonumber\\
&\left.\times \left\{S^c[\Lambda_i^{s}(t_{ij}|\SF_{i, t_{ij}^{-}}^{s})-\Lambda_i^{s}(t_{i,j-1}|\SF_{i, t_{i,j-1}^{-}}^{s})]\right\}^{1-\delta_{ij}^c}\right).
\end{align}
Xu et al.~~\cite{Xuetal2017} use Bayesian methods with diffuse priors to estimate the model parameters. The Metropolis-within-Gibbs algorithm is used to obtained the posterior distributions and then the inference can be carried out using the Markov Chain Monte Carlo (MCMC) samples from the posterior distributions.

\subsection{Prediction for Component Events}\label{sec:prediction}
To make predictions for component events, let $\thetavec$ denote the vector of all the parameters and $\Xvec_i(t_1, t_2)=\{X_i(t);\ t_1<t \leq t_2 \}$ is the covariate information between the time $t_1$ and $t_2$. The prediction for the counts of component events at time $t^{\ast}$ is
\begin{align} \label{eqn:comp.pred}
N_c(t^{\ast}; \thetavec)=&\sum_{i=1}^{n} N_{ic}(t^{\ast}; \thetavec)=\sum_{i=1}^{n} \E_{\Xvec_i(\tau_i, \tau_i+t^{\ast})| \Xvec(\tau_i)}\E_{w_i} \big\{N_{ic}[t^{\ast}, \Xvec_i(\tau_i, \tau_i+t^{\ast}), w_i; \thetavec]\big\}.
\end{align}
Here $N_{ic}(t^{\ast};\thetavec)$ denotes the prediction for unit $i$. Because there is no closed form expression for (\ref{eqn:comp.pred}), the Monte Carlo simulation is used.

\begin{figure}
\begin{center}
\begin{tabular}{cc}
\includegraphics[width=0.42\textwidth]{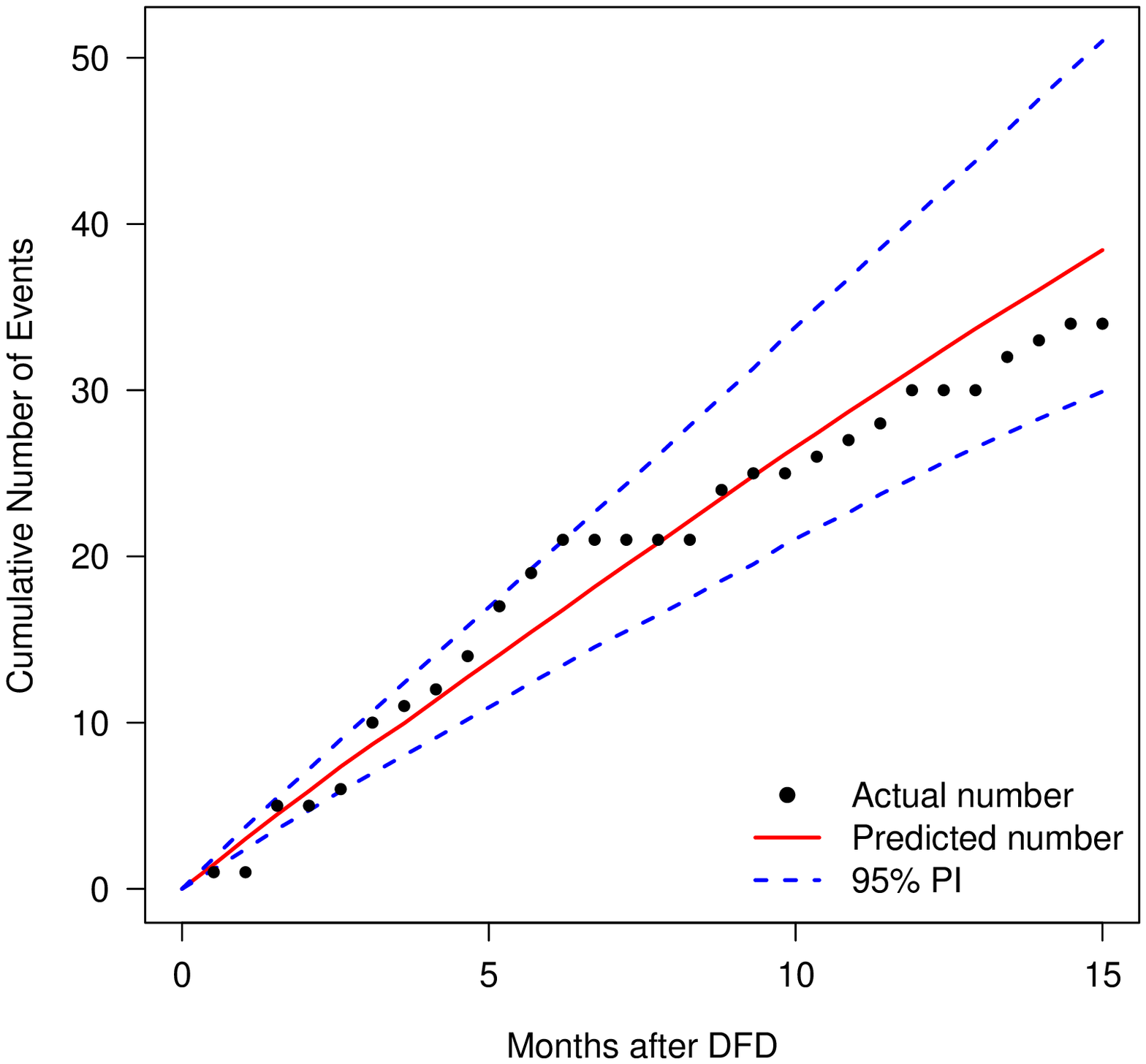} &
\includegraphics[width=0.42\textwidth]{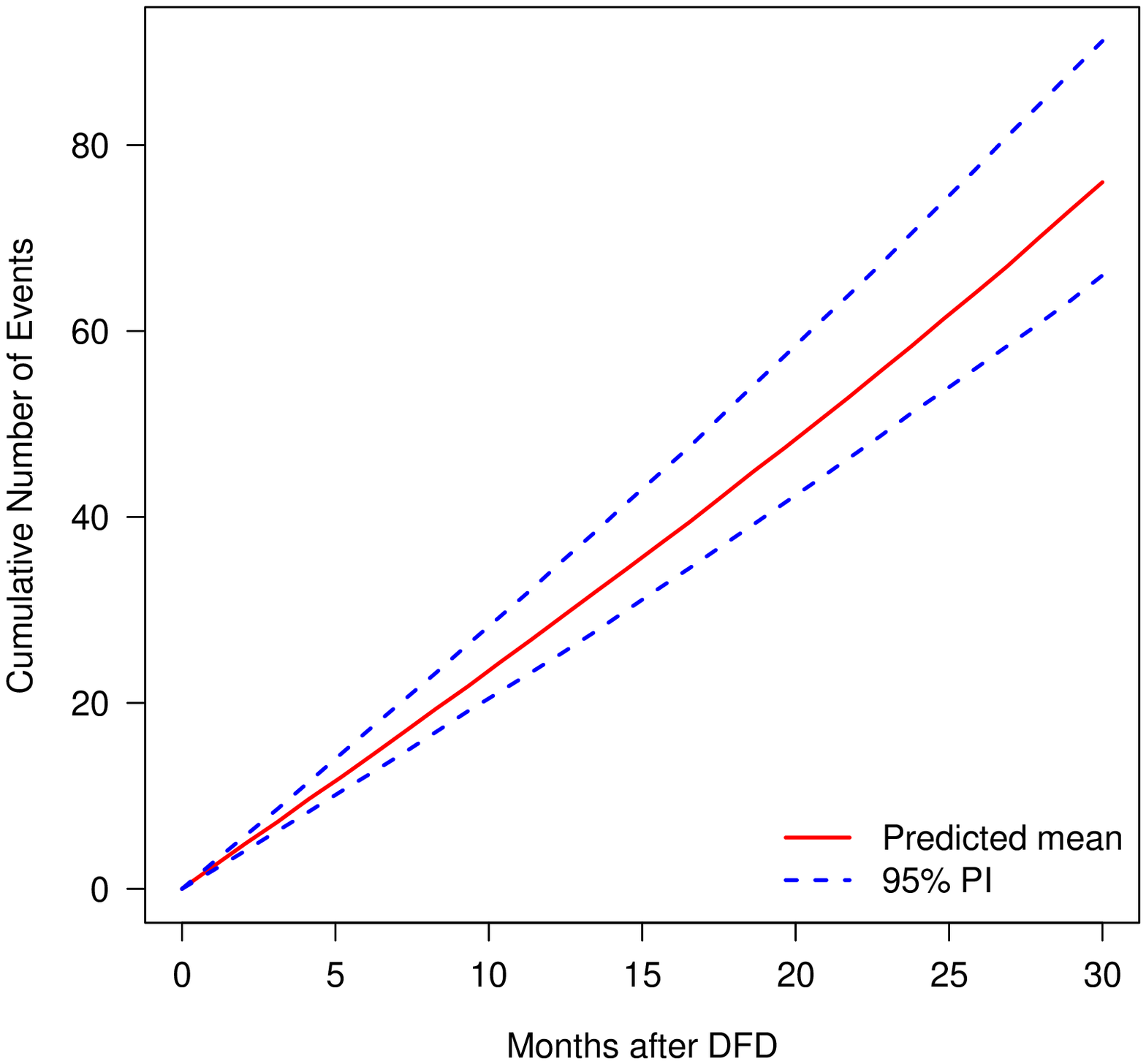}\\
(a) Back test based on an early subset of the data  & (b) Prediction of future events
\end{tabular}
\caption{Plots of the predicted cumulative number of component events for Vehicle b for (a) the last 15 months based on the earlier 95 months data and (b) the future 30 months based on all observed data. \emph{Figure reproduced with permission}.}\label{fig:SA.prediction}
\end{center}
\end{figure}

By fitting the MTRP model to the Vehicle B data using Bayesian estimation, one needs to specify the prior distributions for the unknown parameters. The Weibull distribution was used as renewal functions for $F^c$ and $F^s$. To check the performance of prediction, first the last 15 months of the Vehicle B data were held back and only the first 95 months data were used to estimate the MTRP model and then generate predictions for the last 15 months. Figure~\ref{fig:SA.prediction}(a) shows the prediction of the cumulative component events for the last 15 months based on the earlier data. One can see that the actual observed cumulative numbers of component events are closely located around the predicted values and also well bounded within the pointwise prediction intervals. Figure~\ref{fig:SA.prediction}(b) shows the predicted future events given all the observed data for the next 30 months, which indicates that the total number of component events are expected to range between 62 and 90 with a 95\% confidence level.

\section{Sequential Test Planning of Accelerated Life Tests}\label{sec:seq.alt}
In this section, we briefly introduce the sequential Bayesian design (SBD) for fatigue test experiments described in Lee et al.~\cite{lee2018}.
\subsection{Background and Historical Data}
\begin{figure}
\centering
\begin{tabular}{cc}
\includegraphics[width=.45\textwidth]{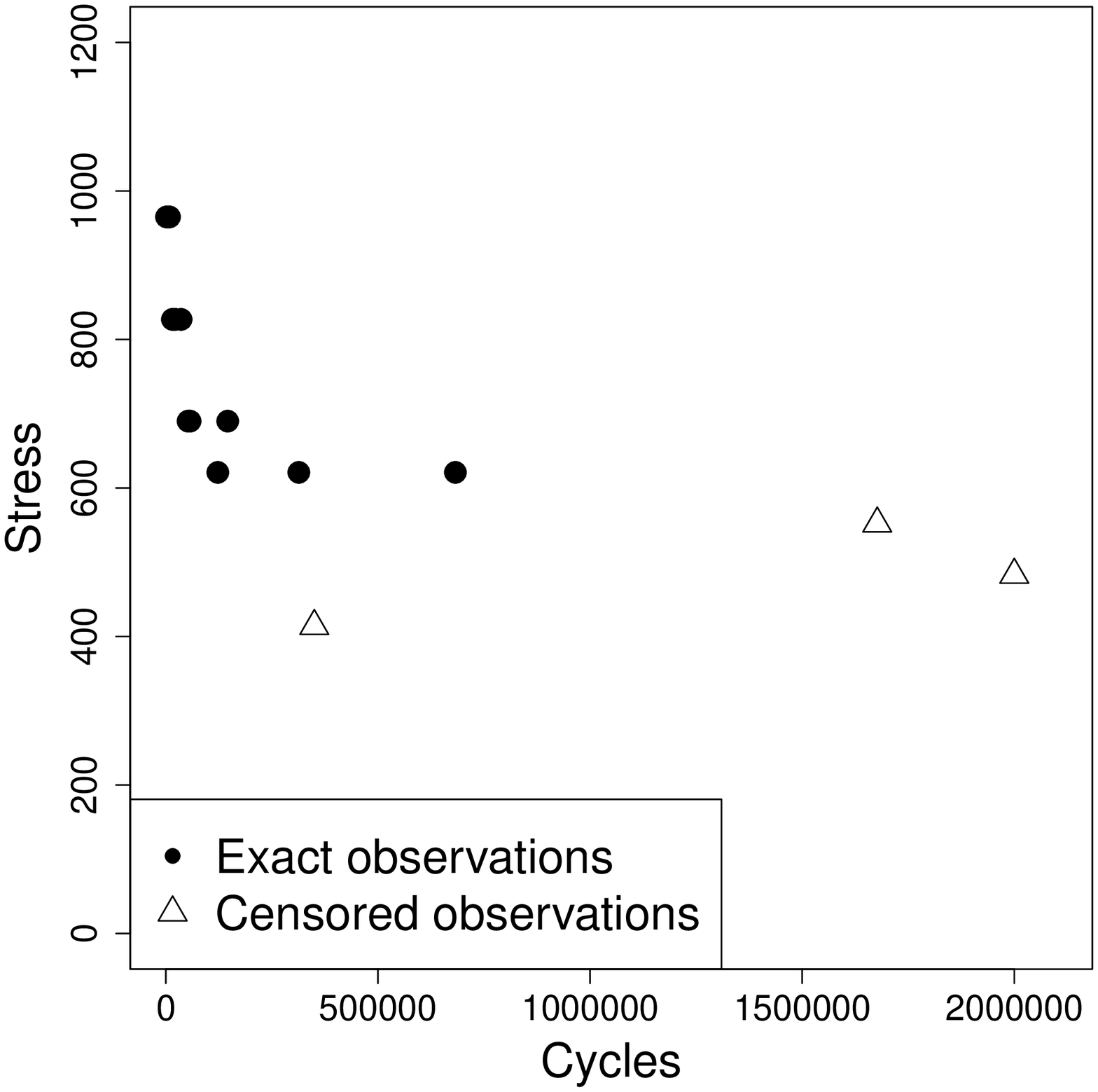} &
\includegraphics[width=.45\textwidth]{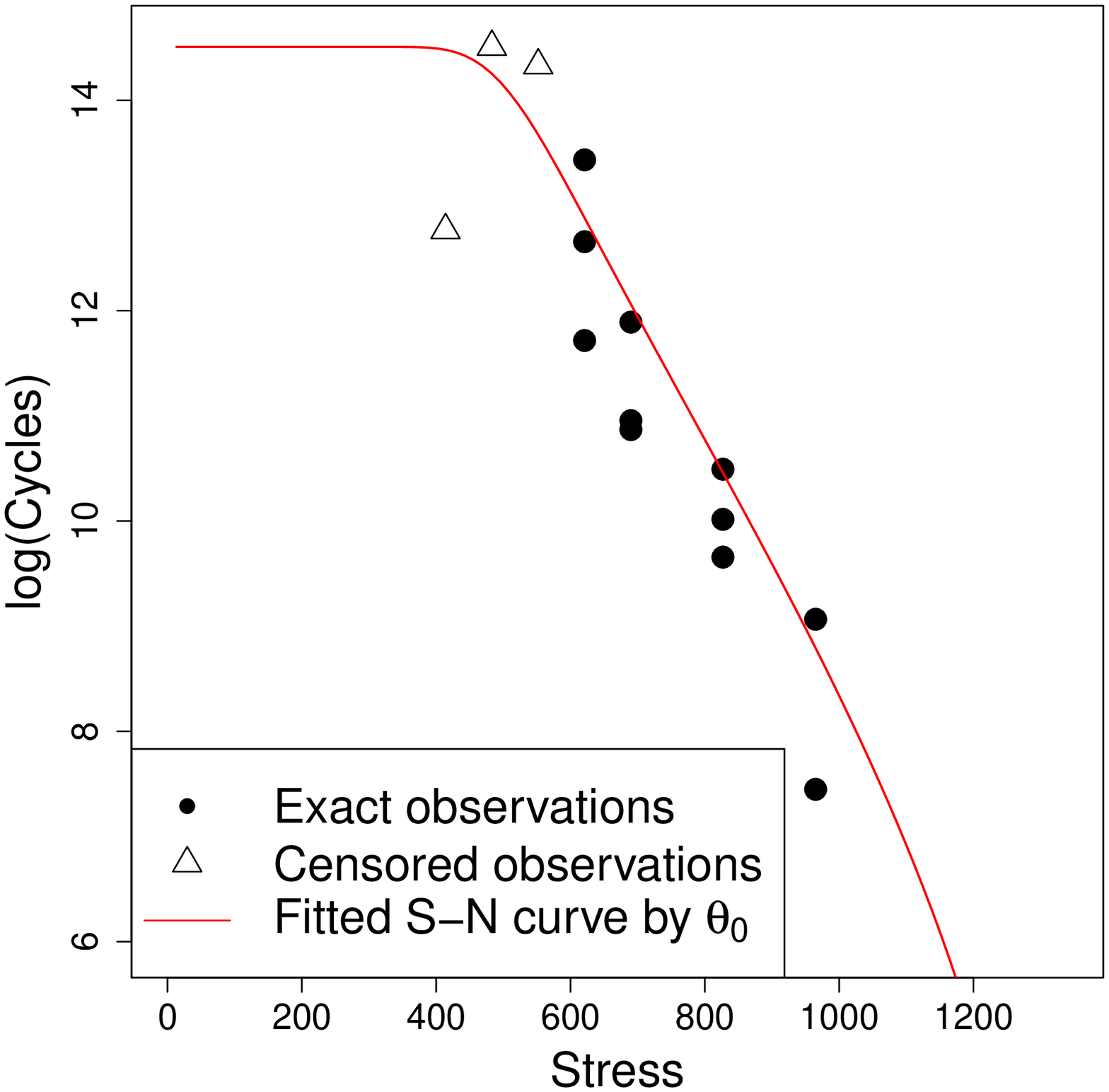}\\
(a) Stress-life relationship & (b) Log of cycles vs. Stress leves\\
\end{tabular}
\caption{The plots show the historical data from a fatigue testing of the glass fiber with the fitted stress-life relationship. \emph{Figure reproduced with permission}.}\label{fig:histdata}
\end{figure}

A sequential Bayesian test planning strategy for the accelerated life tests was proposed by Lee et al.~\cite{lee2018}. The fatigue test for glass fiber, a composite material is considered to illustrate the sequential design strategy. In the test, a tensile/compressive stress $\sigma$ (positive/negative value) is applied to the test unit and the material life is observed under that stress. In this work, 14 observations of E-glass are made including 11 failed and 3 right-censored units. Historical data of the observations are show in Figure~\ref{fig:histdata}. Several other important factors in the test are set as follow. Let $R = \sigma_{m}/\sigma_{M}$ denote the stress ratio, where $\sigma_m$ is the minimum stress and $\sigma_M$ is the maximum stress. The range of $R$ can reveal different test type and it is set at $R = 0.1$ for a tension-tension loading test in this application. The ultimate stress $\sigma_{ult}$, where the material breaks at the first cycle is set to be 1339.67 MPa. The frequency of the cyclic stress testing ($f$) is set at 2 Hz, and the angle ($\alpha$) between the testing direction and material is set at 0.

\subsection{Lifetime Model}

Consider using a log-location-scale distribution to model the cycles-to-failure, $T$. The cdf and pdf are given as
\[F\left(t; \thetavec\right)=\Phi\left[\frac{\log\left(t\right)-\mu}{\nu}\right] \quad \textrm{and} \quad f\left(t; \thetavec\right)=\frac{1}{\nu t}\phi\left[\frac{\log\left(t\right)-\mu}{\nu}\right],\]
where $\Phi(\cdot)$ and $\phi(\cdot)$ are the standard cdf and pdf, respectively. The lognormal and Weibull distributions are the common choices. In the ALT modeling, we assume a constant scale parameter $\nu$ and the location parameter is $\mu=\mu_{\betavec}\left(x\right)$, where $x$ is the stress level and $\betavec$ is the unknown parameter. The following nonlinear model for composite materials proposed in \cite{epaarachchi2003} is used to describe $\mu=\mu_{\betavec}\left(x\right)$ in the form of
\begin{equation}
\mu_{\betavec}(x)=\frac{1}{B}\log\left\{\left(\frac{B}{A}\right)h^B\left(\frac{\ultimatestress}{x}-1\right)
\left(\frac{\ultimatestress}{x}\right)^{\gamma\left(\alpha\right)-1}\left[1-\psi\left(R\right)\right]^{-\gamma\left(\alpha\right)}
+1\right\},
\end{equation}
In the above model, $A$ and $B$ are effects from environment and material, and $\betavec=\left(A,B\right)^\prime$. The function $\psi(R)=1/R$ if $R \geq 1$ and $\psi(R)= R$ if $-\infty < R < 1$, and $\gamma\left(\alpha\right)=1.6-\psi \left\lvert \sin\left(\alpha\right) \right\rvert$. Then $\thetavec = (\betavec^\prime, \nu)^\prime$ denotes the unknown parameter in the ALT modeling.

The lower quantile of the cycles-to-failure distribution is of interest as it contains material life information. The log of the $p$th quantile is
\begin{equation}
\log\left( \xi_{p,u} \right)= \mu_{\betavec}(u) +z_p \nu, \label{eq:log_quantile}
\end{equation}
where $\xi_{p,u}$ is the $p$th quantile at the use condition $u$ and $z_p$ is the $p$th quantile of the standard distribution. Our goal is to propose test planning under multiple use conditions to approximate the real scenarios.  The use stress profile consists of a set of use levels, $\left\{u_1,\cdots,u_K\right\}$, with weights $\left\{w_1,\cdots,w_K\right\}$ and $\sum_{k=1}^K w_k=1$.

Let $\left(x_i,\null t_i,\null \delta_i\right)$ denote the data for the $i$th testing unit, where $x_i$ is the stress level of the accelerating factor and $t_i$ is the observed cycles to failure (or censoring). Let $\delta_i$ be a censoring indicator where $\delta_i$ = 1 if the observation is censored and $\delta_i = 0$ if the observation fails. Then, the log-likelihood function is given by\\
\begin{equation} \label{eq:log_likelihood}
l\left(\thetavec|\xvec_n, \tvec_n, \deltavec_n \right) = \sum_{i=1}^n\left(1-\delta_i\right)\left[\log \phi\left(z_i\right) -\log(t_i) -\log(\nu) \right]+\delta_i \log \left[1-\Phi\left(z_{i}\right)\right],
\end{equation}
where $z_i=\left[\log(t_i)-\mu_{\betavec}(x_i)\right]/\nu$. Let $\widehat{\thetavec}$ be the ML estimates of $\thetavec$ and $\log( \widehat{\xi}_{p,u})$ be the ML estimate of the logarithm of the $p$th quantile at the use level $u$, obtained by substituting $\betavec$ and $\nu$ by $\widehat{\betavec}$ and $\widehat{\nu}$ in (\ref{eq:log_quantile}). Given the use level $u$, the asymptotic variance of $\log( \widehat{\xi}_{p,u})$ is
\[\text{Avar}\left[\log \left(\widehat{\xi}_{p,u}\right)\right]=\boldsymbol{c}^{\prime} \ \Sigma_{\thetavec}(\xvec_n) \  \boldsymbol{c},\]
where $\boldsymbol{c}=\left[\partial \mu_{\betavec}(u)/\partial A, \ \partial\mu_{\betavec}(u)/\partial B, \ z_p\right]'$, $\Sigma_{\thetavec}(\xvec_n)=I_n^{-1}\left( \thetavec \right)$, and $I_n\left( \thetavec \right)$ is the Fisher information matrix based on $n$ observed data. The details for calculating $I_n \left(\thetavec \right)$ can be found in \cite{lee2018}. A weighted version of asymptotic variance can be expressed as
\begin{equation} \label{eq:AVar}
\sum_{k=1}^K w_k \ \text{Avar}\left[ \log \left(\widehat{\xi}_{p,u_k}\right)\right].
\end{equation}
Given $\left\{(u_k, w_k)\right\}_{k=1}^K$, the weighted asymptotic variance only depends on the observed testing levels $x_i$, where $i = 1,\dots,n$. Therefore, the optimum design points should determine $x_1, \ldots, x_n$ to minimize the weighted asymptotic variance in (\ref{eq:AVar}).

\subsection{Test Plan Development}\label{sec:seqdesignstep}
To obtain an accurate prediction from an efficient ALT, the optimum test planning can be determined by minimizing the asymptotic variance in (\ref{eq:AVar}). In the literature, when determining an optimal test plan, it often requires  pre-specifying the values of parameters (known as the planning values). The optimal design based on some chosen planning values of parameters is known as the local $c$-optimality design. However, the planning values are not precisely known a priori for many experiments in practice. Hence, the SBD is useful for improving our understanding  of the unknown parameters as more data becoming available during the experiment, when there is little knowledge or historical data available.

Before the test planning, the stress levels are often standardized to be between 0 and 1, denoted by $q_i = x_i \slash \sigma_{ult}$. In practice, a range of testing levels, $[q_L, q_U]$, is often determined at the very early stage of the test planing, where $q_{L}$ is the lower bound and $q_{U}$ is the upper bound. To design an efficient ALT via the sequential planning, the objective function based on (\ref{eq:AVar}) is chosen as
\begin{equation} \label{eq:Copt}
\varphi \left(q_{\textrm{new}} \right)=\int_{\boldsymbol{\Theta}} \left[ \sum_{k=1}^K w_k\boldsymbol{c_k}^{\prime} \ \Sigma_{\thetavec} \left(\qvec_{\textrm{new}} \right) \boldsymbol{c_k}\right]\pi\left(\thetavec|\qvec_n, \tvec_n, \deltavec_n\right)d\thetavec,
\end{equation}
where $\Sigma_{\thetavec}\left(\qvec_{\textrm{new}}\right)=\left[I_n \left(\thetavec, \qvec_n \right)+ I_1 \left(\thetavec,{q}_{\textrm{new}} \right)\right]^{-1}$, $\qvec_{\textrm{new}}=(\qvec_{n}^\prime, q_{\textrm{new}})^\prime$, $\qvec_n=(q_1, \ldots, q_n)^\prime$, and $\pi\left(\thetavec|\qvec_n, \tvec_n, \deltavec_n\right)$ is the posterior distribution of $\thetavec$. Specifically,
\[\pi\left(\thetavec|\qvec_n, \tvec_n, \deltavec_n\right) \propto  f\left(\tvec_n|\thetavec, \xvec_n, \deltavec_n \right) \pi(\thetavec),\]
where $f\left(\tvec_n|\thetavec, \xvec_n, \deltavec_n \right)$ is the joint pdf of the historical data and $\pi(\thetavec)$ is the prior distribution of $\thetavec$. Then, the optimum $(n+1)$th design point is determined by
\begin{equation} \label{eq:nextpointq}
q_{n+1}^*=\arg \min_{q_{\textrm{new}} \in \left[q_{L}, q_{U}\right]} \varphi \left(q_{\textrm{new}} \right).
\end{equation}

The procedure of the sequential Bayesian design is summarized as follows.
\begin{enumerate}
\item \emph{Specify prior distributions of model parameters.} Specify prior distributions of $A$ and $B$ as $A\sim$ $\textrm{N}(\mu_A, \sigma_A^2)$ and $B\sim$ $\textrm{N}(\mu_B, \sigma_B^2)$, where $\mu_A$, $\sigma_A^2$, $\mu_B$, and $\sigma_B^2$ are the parameters of the normal distribution and set to be known constants. Let $\nu^2 \sim \text{Inverse Gamma}(\kappa, \gamma)$, where $\kappa$ and $\gamma$ can be known from the historical data or experience.
\item \emph{Evaluate the asymptotic variance.} Use the technique of MCMC to approximate (\ref{eq:Copt}). The details of the related algorithms can be found in \cite{lee2018}.
\item \emph{Determine the optimum next testing point $q_{n+1}^*$.} Given a candidate set of design points, their corresponding values of the objective function in (\ref{eq:Copt}) can be evaluated in Step 2. Then, determine the optimum next design point, which has the smallest value of the asymptotic variance.
\item \emph{Obtain the failure data at the level $q_{n+1}^*$.} Under the stress level $q_{n+1}^*$, conduct the experiment and obtain the failure information $(t_{n+1}, \delta_{n+1})$.
\item \emph{Repeat Steps 2 to 4 till the desired number of testing units are obtained.} Add the new lifetime data, $(q_{n+1}^*, t_{n+1}, \delta_{n+1})$, to the historical dataset and repeat Steps 2 to 4 till the desired number of new design points are obtained.
\end{enumerate}

\subsection{Illustration of Test Plans}
For the original data, we can fit the lognormal distribution and the corresponding ML estimates are $\thetavec_0 = \hat{\thetavec} = (0.0157, 0.3188, 0.7259)^\prime$. Before the testing planning, the setup for the sequential Bayesian design is as follows.
\begin{enumerate}
\item \emph{Prior information:} Let $A$ and $B$ be from the normal distributions, where $A \sim \N(0.08$, $0.0008)$ and $B \sim \N(1, 0.0833)$. The prior distribution for $\nu^2$ is  $\text{Inverse Gamma}(4.5, 3)$.
\item \emph{Historical data:} In practical implementation, the sample size at the beginning of testing is limited. Choose the three failed observations at stress levels $\xvec_3 = (621, 690, 965)$ from Figure \ref{fig:histdata} as the historical dataset.
\item \emph{Total size of design points:} Let the sample size of the new design points be $12$.
\item \emph{Design candidate:} The standardized levels of historical data are 0.46, 0.52, and 0.72, and the candidate points are from $q_L = 0.35$ to $q_U = 0.75$ with a 5\% increase.
\end{enumerate}

\begin{figure}
\centering
\begin{tabular}{cc}
\includegraphics[width=0.45\textwidth]{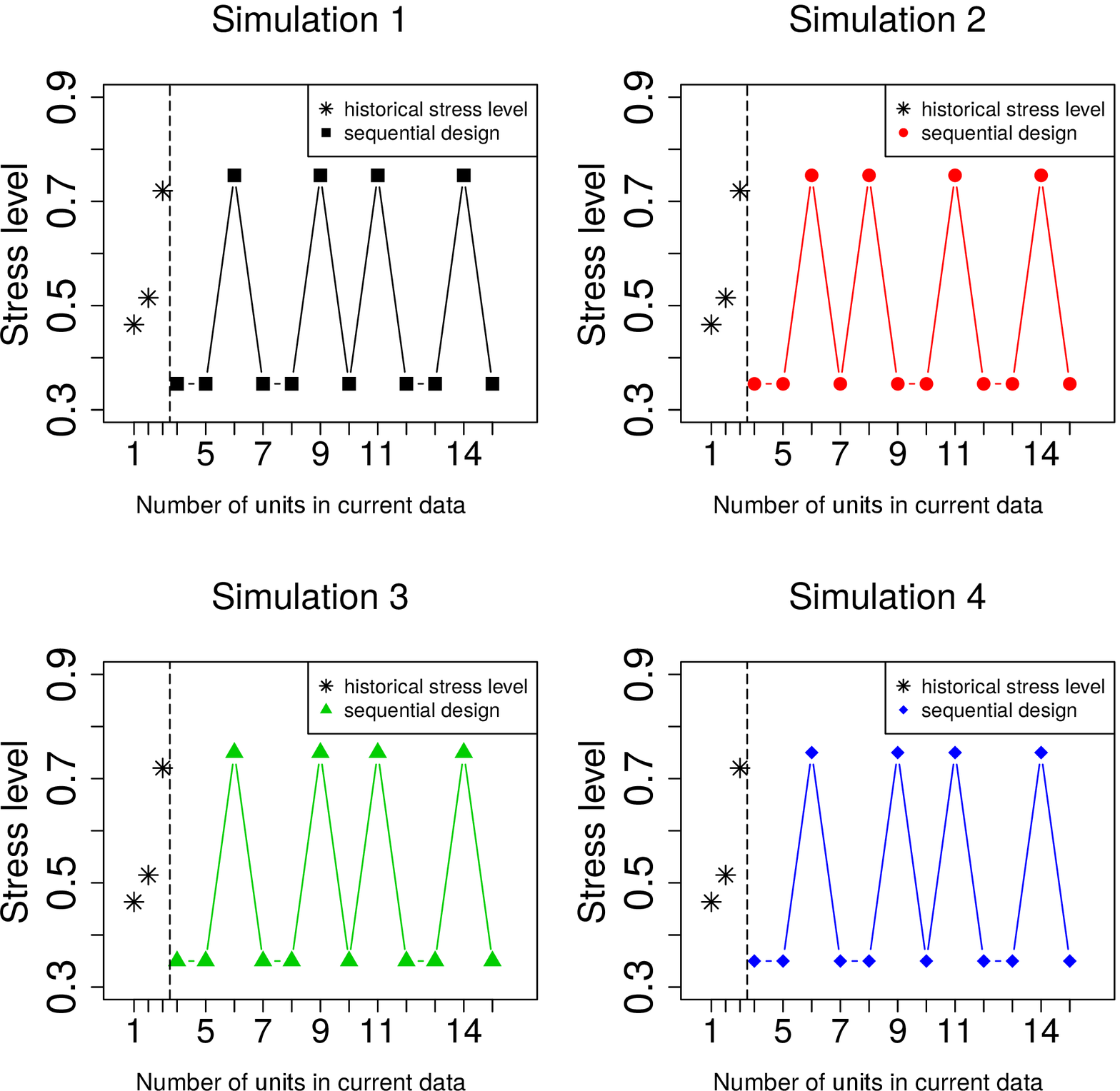} &
\includegraphics[width=0.45\textwidth]{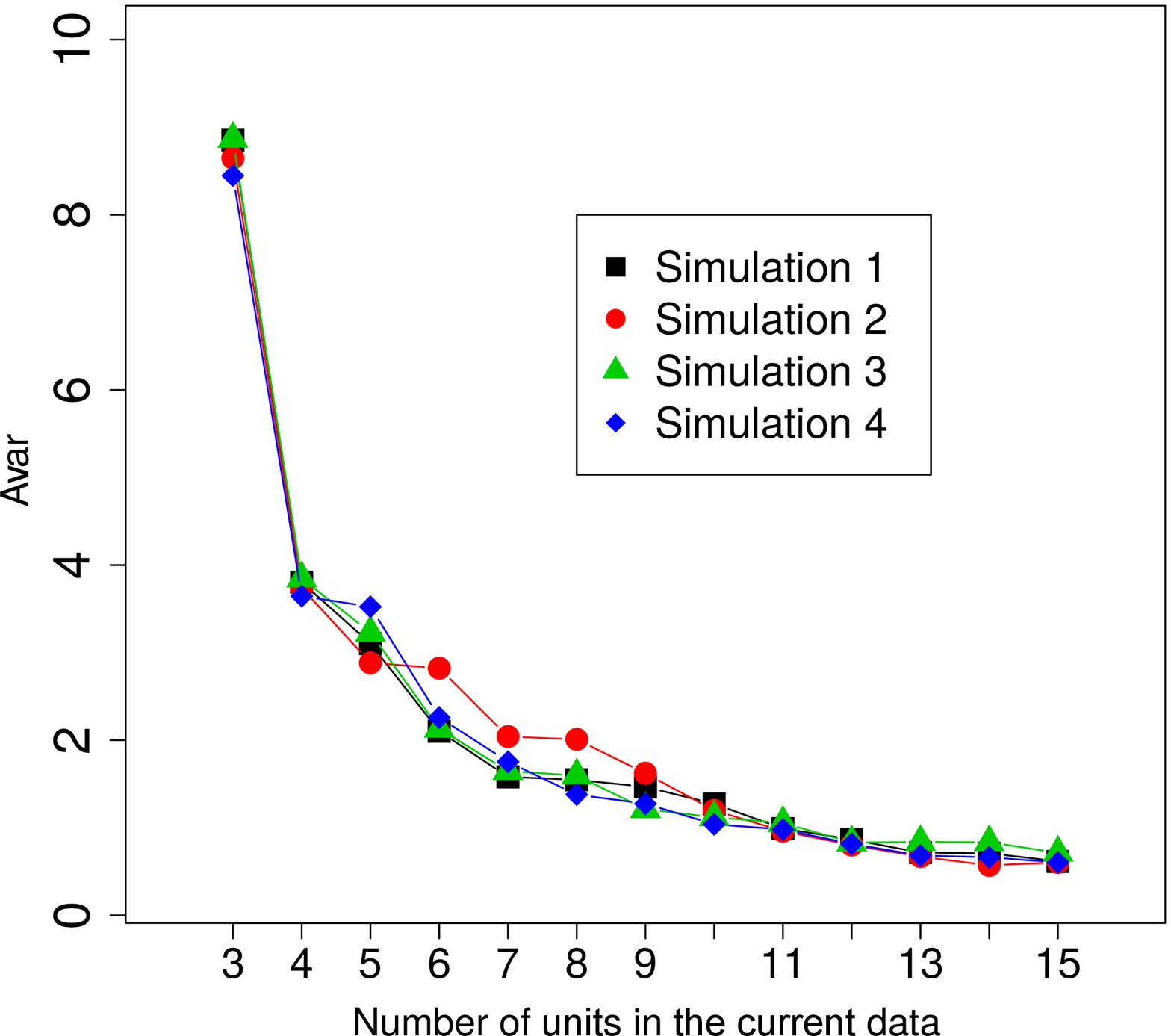}\\
(a) Design points for SBD & (b) Asymptotic variance\\
\end{tabular}
\caption{Plots show the results of the 4 simulation trials including the sequential design points and their corresponding values of asymptotic variance. \emph{Figure reproduced with permission}.}\label{fig:results}
\end{figure}

For the illustrative purpose, assume that the true values of parameters are $\thetavec_0$. When an optimum design point is determined, the new observation is generated from the lognormal distribution with parameter $\thetavec_0$ and the censoring time at $2 \times 10^6$ cycles. Repeat Steps 2 to 4 in Section~\ref{sec:seqdesignstep} till 12 testing locations are obtained. Then, the results of 4 simulation trials are shown in Figure~\ref{fig:results}. It consistently shows that only two stress levels at 0.35 and 0.75 are selected, and 8 and 4 units are allocated to the levels 0.35 and 0.75, respectively. And the resulting asymptotic variances decrease as the size of sequential runs increases.

Using the same historical data, the developed SBD is also compared with the local $c$-optimality design. For the locally c-optimal design, the estimated values of parameters from historical data are usually used as the planning values of the parameters. With only 3 observations available from the historical data, the ML estimates are $\hat{\thetavec}_1 = \left( 0.0005, 0.7429, 0.1658 \right)^\prime$. Hence, the local $c$-optimality design chooses 11 and 1 unit at the testing levels at 0.65 and 0.75, respectively. Now, we compare the performance on the value of the asymptotic variance based on the ML estimates of the final dataset including the 12 new testing observations and 3 historical observations. With 100 simulations, the averages of asymptotic variances for the SBD and the local $c$-optimality designs are $0.6048$ and $4.0337$, respectively. It shows that the SBD is more efficient than the traditional local $c$-optimality design when there is too little historical data available to provide accurate estimates of the model parameters. The proposed SBD can be also applied when there is no historical data but only prior information based on subject matter expertise.


\section{Concluding Remarks}\label{sec:concluding.remark}
In this chapter, we review recent developments on statistical reliability analysis utilizing dynamic covariates and sequential test planning. For time to event data, we introduce a cumulative damage model to account for the effect of dynamic covariates and illustrate the method with the Product D2 application. For degradation data, we present the general path model for incorporating dynamic covariates and illustrate the method with the NIST coating degradation data. We also introduce the MTRP model for recurrent events using dynamic covariates and illustrate it with the Vehicle B data. Regarding to test planning for ALT, we focus on the SBD and illustrate it with the ALT design for polymer composites fatigue testing.

Looking forward, more versatile data become available due to the rapid advance of modern technology, and new statistical methods need to be developed to make use of those new data for improving reliability modeling and prediction. As described in~\cite{HongZhangMeeker2018}, many data types such as spatial data, functional data, image data, and text data, all have great potential to be used for reliability modeling and analysis. New methods that are available in statistics and machine learning can also be transformed and integrated with reliability domain knowledge for reliability analysis, which provides tremendous opportunity in reliability research.



\begin{thebibliography}{10}

\bibitem{meekerescobar1998}
W.~Q. Meeker, L.~A. Escobar:
 \textit{Statistical Methods for Reliability Data}
 (John Wiley \& Sons, Inc., New York 1998)

\bibitem{MeekerHong2014}
W.~Q. Meeker, Y.~Hong:
 Reliability Meets Big Data: Opportunities and Challenges, with Discussion,
 Quality Engineering \textbf{26},
 102--116 (2014)

\bibitem{HongZhangMeeker2018}
Y.~Hong, M.~Zhang, W.~Q. Meeker:
 Big data and reliability applications: The complexity dimension,
 Journal of Quality Technology \textbf{50}(2),
 135--149 (2018)

\bibitem{LawlessCrowderLee2009}
J.~F. Lawless, M.~J. Crowder, K.~A. Lee:
 Analysis of Reliability and Warranty Claims in Products With Age and Usage
  Scales,
 Technometrics \textbf{51},
 14--24 (2009)

\bibitem{Guoetal2009}
H.~Guo, A.~Monteforte, A.~Mettas, D.~Ogden:
 Warranty prediction for products with random stresses and usages. In:
  \textit{IEEE Proceedings Annual Reliability and Maintainability Symposium}
 (IEEE,
 Fort Worth, TX 2009) pp.\,72--77

\bibitem{LuAndersonCook2010}
L.~Lu, C.~M. Anderson-Cook:
 Using Age and Usage for Prediction of Reliability of an Arbitrary System from
  a Finite Population,
 Quality and Reliability Engineering International \textbf{27},
 179--190 (2011)

\bibitem{HongMeeker2010}
Y.~Hong, W.~Q. Meeker:
 Field-failure and Warranty Prediction Based on Auxiliary Use-rate Information,
 Technometrics \textbf{52},
 148--159 (2010)

\bibitem{nelson2001}
W.~Nelson:
 Prediction of field reliability of units, each under differing dynamic
  stresses, from accelerated test data. In: \textit{Handbook of Statistics 20:
  Advances in Reliability}, ed. by N.~Balakrishnan, C.~R. Rao
 (North-Holland, Amsterdam 2001) Chap.\,IX

\bibitem{Voiculescuetal2007}
S.~Voiculescu, F.~Gu\'{e}rin, M.~Barreau, A.~Charki:
 Reliability Estimation in Random Environment: Different Approaches. In:
  \textit{IEEE Proceedings Annual Reliability and Maintainability Symposium}
 (IEEE,
 Orlando, FL 2007) pp.\,202--307

\bibitem{HongMeeker2013}
Y.~Hong, W.~Q. Meeker:
 Field-Failure Predictions Based on Failure-time Data with Dynamic Covariate
  Information,
 Technometrics \textbf{55},
 135--149 (2013)

\bibitem{Whitmore1995}
G.~A. Whitmore:
 Estimation degradation by a \protect{Wiener} diffusion process subject to
  measurement error,
 Lifetime Data Analysis \textbf{1},
 307--319 (1995)

\bibitem{DoksumHoyland1992}
K.~A. Doksum, A.~H\'{o}yland:
 Models for variable-stress accelerated life testing experiments based on
  \protect{Wiener} processes and the inverse \protect{Gaussian} distribution,
 Technometrics \textbf{34},
 74--82 (1992)

\bibitem{Wang2010}
X.~Wang:
 Wiener processes with random effects for degradation data,
 Journal of Multivariate Analysis \textbf{101},
 340--351 (2010)

\bibitem{LawlessCrowder2004}
J.~F. Lawless, M.~Crowder:
 Covariates and random effects in a gamma process model with application to
  degradation and failure,
 Lifetime Data Analysis \textbf{10},
 213--227 (2004)

\bibitem{WangXu2010}
X.~Wang, D.~Xu:
 An Inverse \protect{Gaussian} Process Model for Degradation Data,
 Technometrics \textbf{52},
 188--197 (2010)

\bibitem{YeChen2014}
Z.-S. Ye, N.~Chen:
 The Inverse {Gaussian} Process as a Degradation Model,
 Technometrics \textbf{56},
 302--311 (2014)

\bibitem{LuMeeker1993}
C.~J. Lu, W.~Q. Meeker:
 Using degradation measures to estimate a time-to-failure distribution,
 Technometrics \textbf{34},
 161--174 (1993)

\bibitem{MeekerEscobarLu1998}
W.~Q. Meeker, L.~A. Escobar, C.~J. Lu:
 Accelerated degradation tests: modeling and analysis,
 Technometrics \textbf{40},
 89--99 (1998)

\bibitem{BagdonaviciusNikulin2001}
V.~Bagdonavi\v{c}ius, M.~S. Nikulin:
 Estimation in degradation models with explanatory variables,
 Lifetime Data Analysis \textbf{7},
 85--103 (2001)

\bibitem{BaeKuoKvam2007}
S.~J. Bae, W.~Kuo, P.~H. Kvam:
 Degradation models and implied lifetime distributions,
 Reliability Engineering and System Safety \textbf{92},
 601--608 (2007)

\bibitem{Duanetal2017}
Y.~Duan, Y.~Hong, W.~Meeker, D.~Stanley, X.~Gu:
 Photodegradation Modeling Based on Laboratory Accelerated Test Data and
  Predictions Under Outdoor Weathering for Polymeric Materials,
 The Annals of Applied Statistics \textbf{11},
 2052--2079 (2017)

\bibitem{EscobarMeekerKuglerKramer2003}
L.~A. Escobar, W.~Q. Meeker, D.~L. Kugler, L.~L. Kramer:
 Accelerated Destructive Degradation Tests: Data, Models, and Analysis. In:
  \textit{Mathematical and Statistical Methods in Reliability}, ed. by B.~H.
  Lindqvist, K.~A. Doksum
 (World Scientific Publishing Company, Singapore 2003)

\bibitem{Xieetal2018}
Y.~Xie, C.~B. King, Y.~Hong, Q.~Yang:
 Semi-parametric models for accelerated destructive degradation test data
  analysis,
 Technometrics \textbf{60},
 222--234 (2018)

\bibitem{Dingetal2019}
Y.~Ding, Q.~Yang, C.~B. King, Y.~Hong:
 A General Accelerated Destructive Degradation Testing Model for Reliability
  Analysis,
 IEEE Transactions on Reliability, DOI: 10.1109/TR.2018.2883983
 (2019)

\bibitem{HongDuanetal2015}
Y.~Hong, Y.~Duan, W.~Q. Meeker, D.~L. Stanley, X.~Gu:
 Statistical Methods for Degradation Data with Dynamic Covariates Information
  and an Application to Outdoor Weathering Data,
 Technometrics \textbf{57},
 180--193 (2015)

\bibitem{XuHongJin2015}
Z.~Xu, Y.~Hong, R.~Jin:
 Nonlinear General Path Models for Degradation Data with Dynamic Covariates,
 Applied Stochastic Models in Business and Industry \textbf{32},
 153--167 (2016)

\bibitem{ZhaoLiu2003}
R.~Zhao, B.~Liu:
 Renewal process with fuzzy interarrival times and rewards,
 International Journal of Uncertainty, Fuzziness and Knowledge-Based Systems
  \textbf{11},
 573--586 (2003)

\bibitem{HongLiOsborn2015}
Y.~Hong, M.~Li, B.~Osborn:
 System unavailability analysis based on window-observed recurrent event data,
 Applied Stochastic Models in Business and Industry \textbf{31},
 122--136 (2015)

\bibitem{Kijima1989}
M.~Kijima:
 Some results for repairable systems with general repair,
 Journal of Applied Probability \textbf{26},
 89--102 (1989)

\bibitem{PhamandWang1996}
H.~Wang, H.~Pham:
 A quasi renewal process and its applications in imperfect maintenance,
 International Journal of Systems Science \textbf{27},
 1055--1062 (1996)

\bibitem{DoyenandGaudoin2004}
L.~Doyen, O.~Gaudoin:
 Classes of imperfect repair models based on reduction of failure intensity or
  virtual age,
 Reliability Engineering and System Safety \textbf{84},
 45--56 (2004)

\bibitem{LindqvistElvebakkHeggland2003}
B.~Lindqvist, G.~Elvebakk, K.~Heggland:
 The trend-renewal process for statistical analysis of repairable systems,
 Technometrics \textbf{45},
 31--44 (2003)

\bibitem{YangShi2012}
Q.~Yang, Y.~Hong, Y.~Chen, J.~Shi:
 Failure Profile Analysis of Complex Repairable Systems with Multiple Failure
  Modes,
 IEEE Transactions on Reliability \textbf{61},
 180--191 (2012)

\bibitem{PietznerWienke2013}
D.~Pietzner, A.~Wienke:
 The trend-renewal process: a useful model for medical recurrence data,
 Statistics in Medicine \textbf{32},
 142--152 (2013)

\bibitem{Yangetal2017}
Q.~Yang, Y.~Hong, N.~Zhang, J.~Li:
 A copula-based trend-renewal process model for analysis of repairable systems
  with multitype failures,
 IEEE Transactions on Reliability \textbf{66}(3),
 590--602 (2017)

\bibitem{Xuetal2017}
Z.~Xu, Y.~Hong, W.~Q. Meeker, B.~E. Osborn, K.~Illouz:
 A Multi-level Trend-Renewal Process for Modeling Systems with Recurrence Data,
 Technometrics \textbf{59},
 225--236 (2017)

\bibitem{meeker1984}
W.~Q. Meeker:
 A comparison of accelerated life test plans for {Weibull} and lognormal
  distributions and type {I} censoring,
 Technometrics \textbf{26}(2),
 157--171 (1984)

\bibitem{nelson1990}
W.~Nelson:
 \textit{Accelerated Testing: Statistical Models, Test Plans, and Data
  Analyses, (Republished in a paperback in Wiley Series in Probability and
  Statistics, 2004)}
 (John Wiley \& Sons, New York 1990)

\bibitem{zhang2005}
Y.~Zhang, W.~Q. Meeker:
 Bayesian life test planning for the {Weibull} distribution with given shape
  parameter,
 Metrika \textbf{61}(3),
 237--249 (2005)

\bibitem{zhang2006}
Y.~Zhang, W.~Q. Meeker:
 Bayesian methods for planning accelerated life tests,
 Technometrics \textbf{48}(1),
 49--60 (2006)

\bibitem{Hongetal2015}
Y.~Hong, C.~B. King, Y.~Zhang, W.~Q. Meeker:
 Bayesian life test planning for log-location-scale family of distributions,
 Journal of Quality Technology \textbf{47},
 336--350 (2015)

\bibitem{Kingetal2016}
C.~King, Y.~Hong, S.~P. Dehart, P.~A. Defeo, R.~Pan:
 Planning Fatigue Tests for Polymer Composites,
 Journal of Quality Technology \textbf{48},
 227--245 (2016)

\bibitem{lee2018}
I.-C. Lee, Y.~Hong, S.-T. Tseng, T.~Dasgupta:
 Sequential Bayesian Design for Accelerated Life Tests,
 Technometrics \textbf{60}(4),
 472--483 (2018)

\bibitem{LuLeeHong2019}
L.~Lu, I.~Lee, Y.~Hong:
 Bayesian Sequential Design Based on Dual Objectives for Accelerated Life
  Tests,
 arXiv:1812.00055
 (2019)

\bibitem{bagnik2001}
V.~Bagdonavi\v{c}ius, M.~S. Nikulin:
 \textit{Accelerated Life Models: Modeling and Statistical Analysis}
 (Chapman \& Hall/CRC, Boca Raton, FL 2001)

\bibitem{LawlessFredette2005}
J.~F. Lawless, M.~Fredette:
 Frequentist prediction intervals and predictive distributions,
 Biometrika \textbf{92},
 529--542 (2005)

\bibitem{Hong2013}
Y.~Hong:
 On Computing the Distribution Function for the {Poisson} Bionomial
  Distribution,
 Computational Statistics and Data Analysis \textbf{59},
 41--51 (2013)

\bibitem{Meyer2008}
M.~C. Meyer:
 Inference using shape-restricted regression splines,
 The Annals of Applied Statistics \textbf{2},
 1013--1033 (2008)

\bibitem{FraserMassam1989}
D.~A.~S. Fraser, H.~Massam:
 A Mixed Primal-Dual Bases Algorithm for Regression under Inequality
  Constraints. \protect{Application} to Concave Regression,
 Scandinavian Journal of Statistics \textbf{16},
 65--74 (1989)

\bibitem{CarppenterGoldsteinrasbash2003}
J.~R. Carpenter, H.~Goldstein, J.~Rasbash:
 A novel bootstrap procedure for assessing the relationship between class size
  and achievement,
 Applied Statistics \textbf{52},
 431--443 (2003)

\bibitem{Lutkepohl2005}
H.~L\"{u}tkepohl:
 \textit{New Introduction to Multiple Time Series Analysis}, Second edn.
 (Springer-Verlag, Berlin 2005)

\bibitem{epaarachchi2003}
J.~A. Epaarachchi, P.~D. Clausen:
 An empirical model for fatigue behavior prediction of glass fibre-reinforced
  plastic composites for various stress ratios and test frequencies,
 Composites Part A: Applied science and manufacturing \textbf{34}(4),
 313--326 (2003)

\end{thebibliography}

\end{document}